%% file: main.tex
\documentclass[acmtog]{acmart}

\graphicspath{{figures/}}
\input{header.tex}

\title{Power-Linear Polar Directional Fields}
\author{Jiabao Brad Wang}
\affiliation{\institution{University of Edinburgh}
\country{United Kingdom}
}
\email{bradwangw@gmail.com}
\author{Amir Vaxman}
\affiliation{\institution{University of Edinburgh}
\country{United Kingdom}
}
\email{avaxman@inf.ed.ac.uk}
\acmSubmissionID{128}

\begin{document}


\begin{abstract}
We introduce a novel method for directional-field design on meshes, enabling users to specify singularities at any location on a mesh. Our method uses a piecewise power-linear representation for phase and scale, offering precise control over field topology. The resulting fields are smooth and accommodate any singularity index and field symmetry. With this representation, we mitigate the artifacts caused by coarse or uneven meshes. We showcase our approach on meshes with diverse topologies and triangle qualities.
\end{abstract}

\begin{teaserfigure}
\centering
\includegraphics[width=0.65\textwidth]{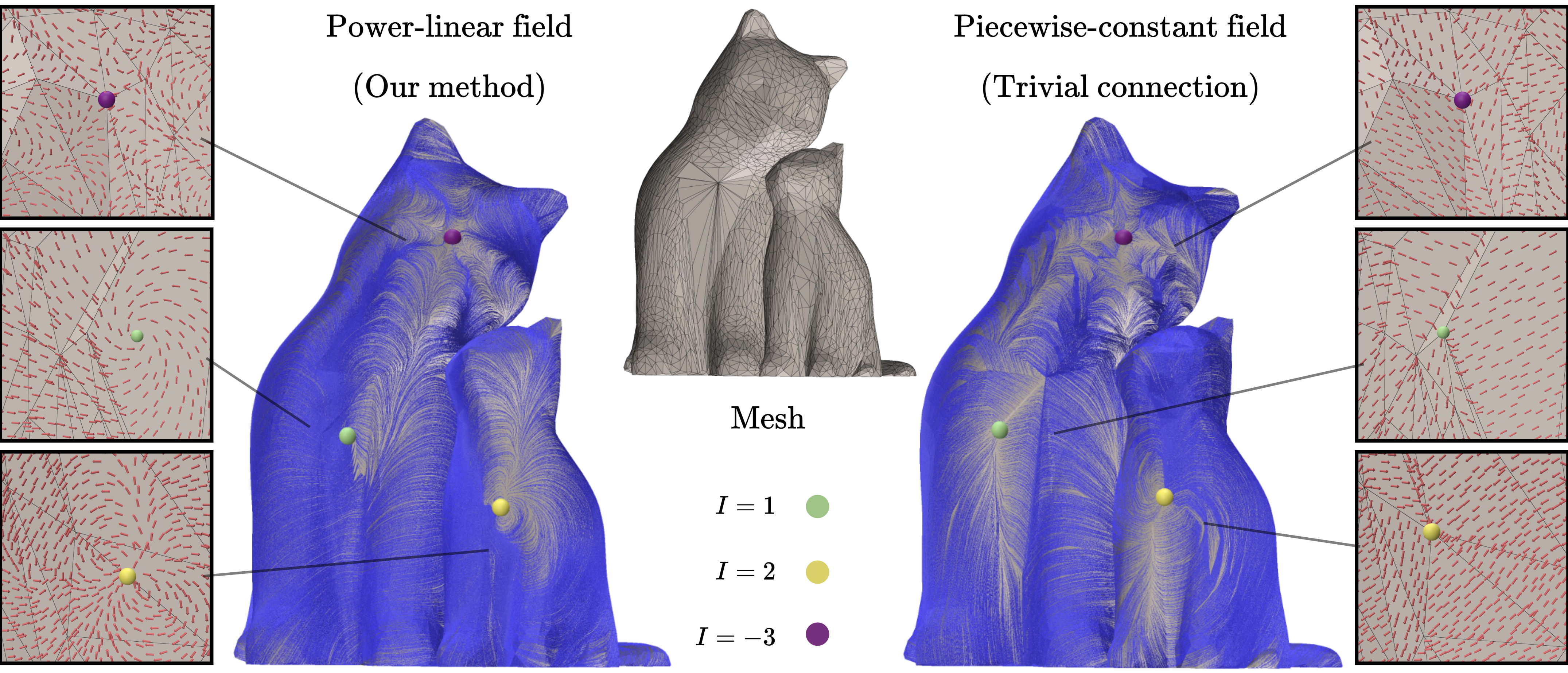}
    \caption{We compute rotationally-smooth directional fields by prescribing their topology explicitly. Our fields (left) are piecewise power-linear, and therefore allow for singularities of any index defined at any point on the mesh, including on edges and inside faces. This generalizes common piecewise-constant approaches (right). Our approach results in smoother fields in comparison, even for uneven and coarse meshes.}
    \label{fig:teaser}
\end{teaserfigure}

\maketitle

\input{sections/intro.tex}
\input{sections/relatedwork.tex}
\input{sections/piecewise-linear-polar-fields}
\input{sections/bevelled-meshes}
\input{sections/results.tex}
\input{sections/conclusion.tex}

\begin{acks}
The authors would like to thank the anonymous reviewers for their valuable feedback. Special thanks go to Justin Solomon and Albert Chern for helpful discussions.
\end{acks}

\bibliographystyle{ACM-Reference-Format}
\bibliography{bibliography}

\begin{figure*}
\centering
\includegraphics[height=0.27\textheight]{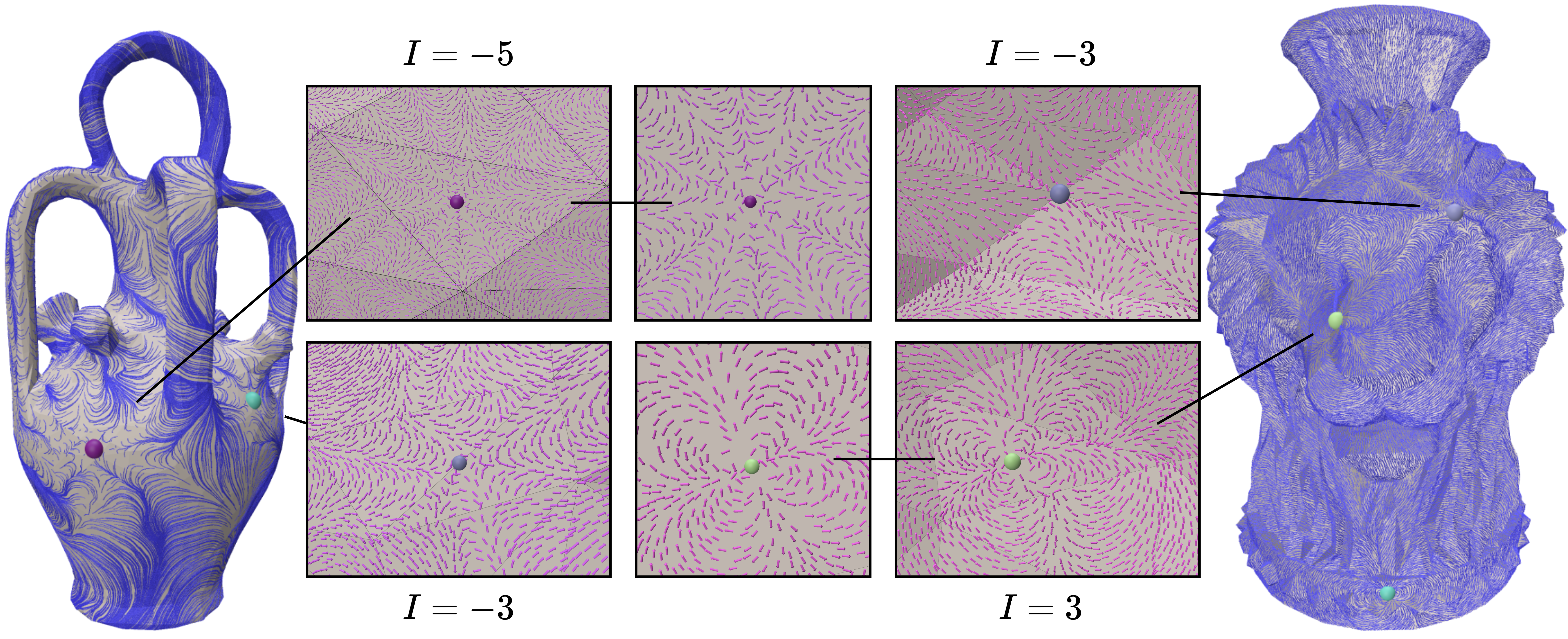}
\caption{Higher-order singularities are well-captured by our method, on vertices, edges, or faces, due to our power-linear representation.}
\label{fig:high-index}
\end{figure*}

\begin{figure*}
\centering
\includegraphics[height=0.25\textheight]{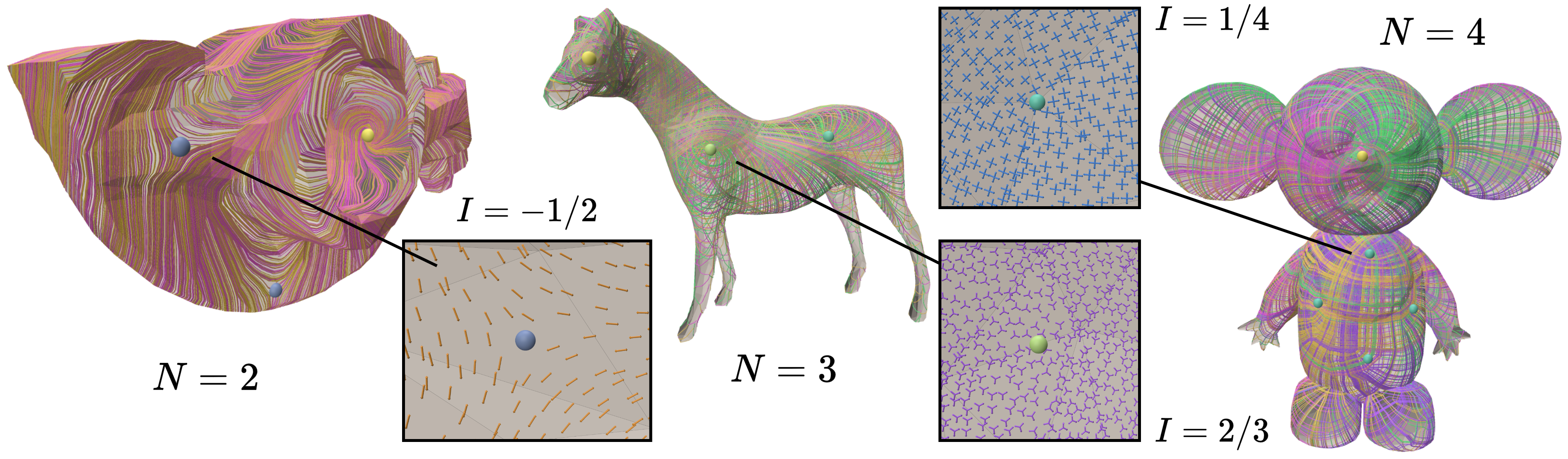}
\caption{$N$-fields with vertex, edge, and face singularities of different indices.}
\label{fig:power-fields}
\end{figure*}

\begin{figure*}
\centering
\includegraphics[height=0.25\textheight]{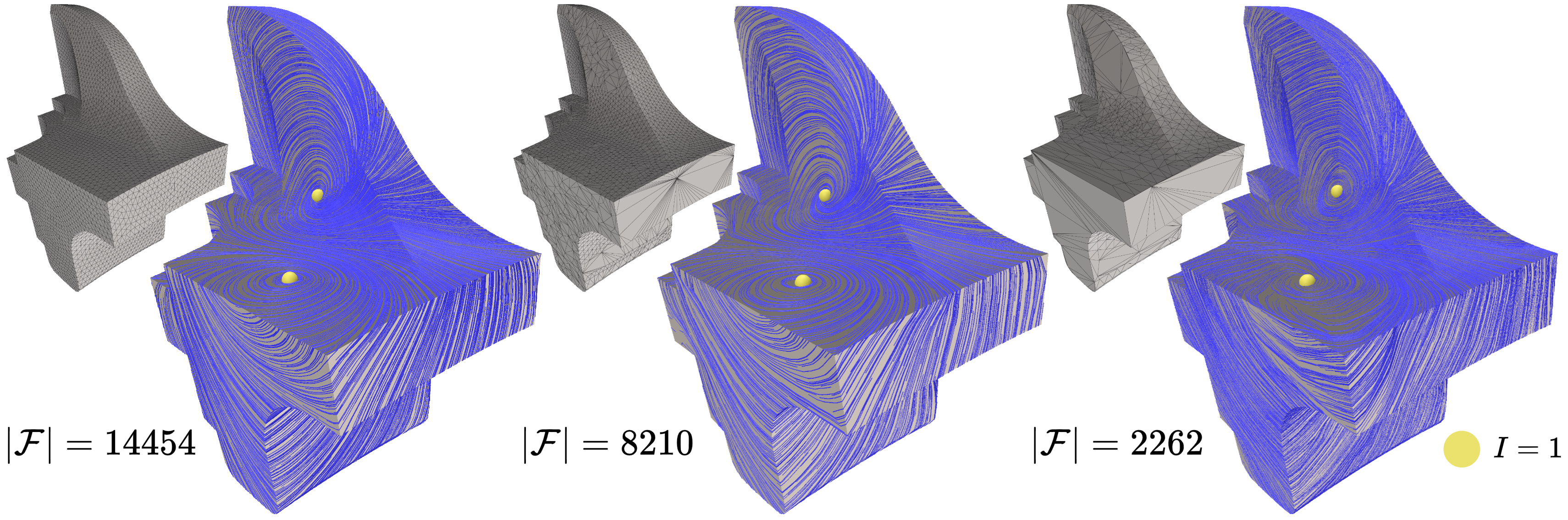}
\caption{Our method produces similar fields for the same geometry, despite considerable differences in mesh quality.}
\label{fig:mesh-quality}
\end{figure*}


\begin{figure*}
\centering
\includegraphics[width=.9\textwidth]{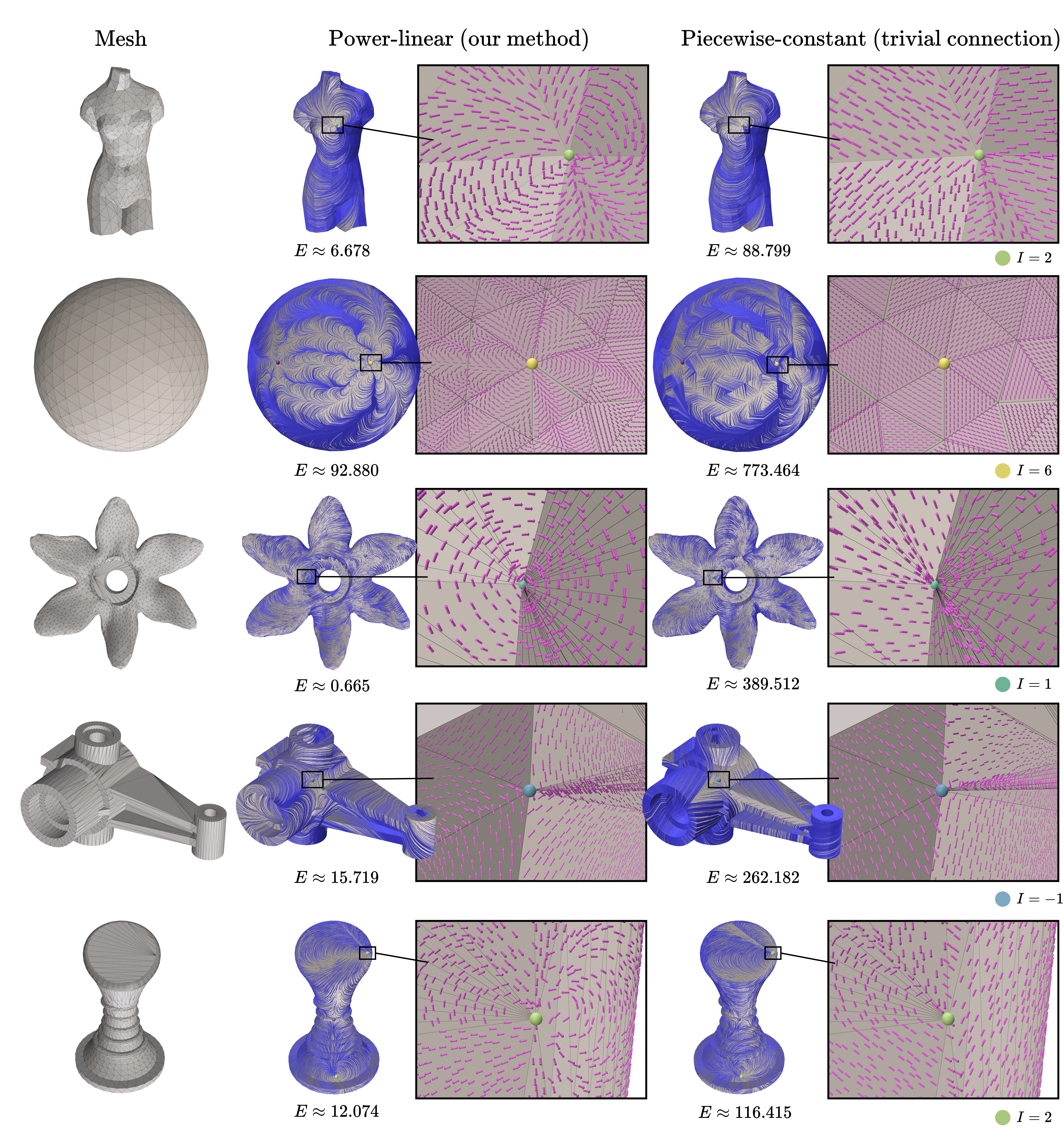}
\caption{Comparison to trivial connections. It is apparent that our method, due to being of a higher order, accommodates higher-order singularities and uneven meshes better than a piecewise-constant representation.}
\label{fig:comparison-trivial-connection}
\end{figure*}

\pagebreak
\appendix
\input{appendix.tex}


\end{document}

%% file: header.tex
\usepackage{booktabs} 
\usepackage{units} 

\usepackage[linesnumbered,ruled,vlined]{algorithm2e}

\SetAlFnt{\normalsize}
\SetAlCapFnt{\normalsize}
\SetAlCapNameFnt{\normalsize}
\SetAlCapHSkip{0pt}
\usepackage{hhline}
\usepackage{bbold} 
\usepackage{amsfonts} 
\usepackage{amsmath} 
\usepackage{pgfplots}
\usepackage{pgf,tikz}
\usepackage{tkz-euclide}
\usepackage{bm}
\usepackage{wrapfig}
\usepackage{multirow}
\usepackage{mathtools}
\usepackage{enumerate}
\usepackage{enumitem}
\usepackage{verbatim}
\usepackage{xcolor}

\usepackage{dsfont}

\newcommand{\M}{\mathcal{M}}

\newcommand{\V}{\mathcal{V}}
\newcommand{\E}{\mathcal{E}}

\newcommand{\F}{\mathcal{F}}

\usepackage{subcaption}

\DeclareMathOperator*{\argmin}{arg\,min}

\usepackage{soul}

\usepackage{listings}
\usepackage{color, colortbl} 
\definecolor{mygreen}{RGB}{28,172,0} 
\definecolor{mylilas}{RGB}{170,55,241}
\definecolor{ffzzcc}{rgb}{1,0.6,0.8}
\definecolor{mytbcol}{RGB}{175,227,246}
\definecolor{mypink}{RGB}{254, 197, 187}
\definecolor{tabyellow}{HTML}{7AAAF2} 

\usepackage{xurl}

\theoremstyle{definition}

\usepackage{lipsum}

\DeclareFontFamily{U}{mathx}{\hyphenchar\font45}
\DeclareFontShape{U}{mathx}{m}{n}{<-> mathx10}{}
\DeclareSymbolFont{mathx}{U}{mathx}{m}{n}

\usepackage{tipa}
\UndeclareTextCommand{\!}{T3}
\DeclareTextCommand{\tipaEXCLAM}{T3}{}
\DeclareRobustCommand{\!}{%
  \ifmmode\mskip-\thinmuskip\else\expandafter\tipaEXCLAM\fi
}

\usepackage{booktabs}       
\usepackage{mathrsfs}
\usetikzlibrary{arrows}
\usetikzlibrary{matrix}
\usetikzlibrary{positioning,calc,fadings}
\usetikzlibrary{backgrounds, fit}
\pgfplotsset{compat=1.14}
\usepackage{pgfplotstable}

\definecolor{myblue}{RGB}{23, 195, 178}
\definecolor{myyellow}{RGB}{255, 186, 8}
\definecolor{mygray}{rgb}{0.5,0.5,0.5}
\definecolor{myred}{RGB}{254, 109, 115}

\definecolor{colblue}{HTML}{7fc8f8}
\definecolor{colyellow}{HTML}{ffe45e}

\definecolor{colvu}{HTML}{99d98c}
\definecolor{colvp}{HTML}{7fc8f8}
\definecolor{coleu}{HTML}{ff8fab}
\definecolor{colep}{HTML}{ffd000} 
\definecolor{coled}{HTML}{99d98c}

\newif\ifdraft

\drafttrue

\ifdraft

\newcommand{\todo}[1]{}

\newcommand{\bld}[1]{\boldsymbol{#1}}
\newcommand{\vectheta}{\bld{\theta}}
\newcommand{\vecsigma}{\bld{\sigma}}

\fi

\usepackage{accsupp}

%% file: sections/intro.tex
\section{Introduction}
\label{sec:intro}
The design and analysis of directional fields are at the core of many pipelines in geometry processing, including seamless parameterization for meshing (e.g.,~\cite{Bommes:2009, Bommes:2013, Shen:2022}), architectural and industrial design~\cite{Verhoeven:2022, segall:2024}, procedural generation of surface details~\cite{Knoppel:2015, chen:2023}, discrete differential geometry~\cite{Sharp:2019, palmer:2024}, and more. Directional-field algorithms are broadly divided into two categories: \emph{Cartesian methods} represent fields with Cartesian coordinates $(x,y,(z))$ on surfaces and in volumes, and allow alignment to features (e.g., principal directions), and automatic computation of singularities. Conversely, \emph{polar methods} work with phases $\psi$ and scales $\sigma$ for $\sigma e^{i\psi}$, and thus allow explicit control of field topology. They also have interesting theoretical links with optimal transport~\cite{Solomon:2019}, and conformal mapping~\cite{weber:2010}.
Our method is of the polar type, which has been far less explored in the literature. Polar methods usually rely on low-order or piecewise-constant fields. As a result, they restrict the positioning of singularities, and result in non-smooth solutions (Figs.~\ref{fig:teaser} and~\ref{fig:comparison-trivial-connection}). Such artifacts are exacerbated by bad meshes and with higher-index singularities.

Recent methods~\cite{boksebeld_high-order_2022, deGoes:2016sec, Manded:2020} explored the merits of higher-order fields in the Cartesian setting, and our work explores higher-order fields in the polar setting for the first time. Phases and scales are not well-defined around singularities, and thus we cannot simply use polynomial finite-element spaces for them, as is common for Cartesian methods. We instead work with phases and scales of piecewise-continuous \emph{power-linear} fields of the form:
\begin{equation*}
\bld{u}(z) = (az+b\overline{z}+c)^I,\ \ a,b,c,z \in \mathbb{C},\ I \in \mathbb{Z} \setminus \{0\}.
\end{equation*}
This representation comprises a rich space of fields, including piecewise holomorphic and anti-holomorphic fields. It is an order higher than piecewise constant fields, which allows to mitigate the artifacts of low-order representations considerably. It moreover allows a user to prescribe singularities \emph{anywhere} on a triangle mesh, be its faces, edges, or vertices. Our contributions are:
\begin{itemize}[itemsep=1pt, parsep=0pt, left=0pt, topsep=0pt]
    \item We define piecewise linear polar fields (Sec.~\ref{sec:piecewise-linear-polar-fields}) and extend them to power-linear fields that represent higher-index singularities and $N$-symmetric fields (Sec.~\ref{subsec:power-linear fields}). 
    \item We introduce a novel \emph{beveled} representation of a triangle mesh that uniformizes piecewise continuity (Sec.~\ref{sec:fields-on-bevelled-meshes}).
    \item We optimize for the field phase and scale by solving a linear and a convex system (Sec.~\ref{subsec:optimization}).
\end{itemize}

%% file: sections/relatedwork.tex
\section{Related Work}
\label{sec:related-work}
Directional field design and synthesis on $2$-manifold surfaces have been studied extensively, and summarized in two surveys~\cite{Vaxman:2016,deGoes:2016vf}; we hereby only summarize the works related to our novelty and methodology.
\paragraph{Cartesian methods} Recent methods like ~\cite{Verhoeven:2022, segall:2024, palmer:2024} and~\cite{Vekhter:2019, Nabizadeh:2022, Sharp:2019})
work with Cartesian representations, for their simplicity and beneficial properties. This includes methods for representing tensors (e.g.,~\cite{Zhang:2007, Couplet:2024}). Perhaps the most important of such properties is integrability~\cite{Diamanti:2015, Sageman:2019}, leading to low-error seamless parameterizations. Their shortcoming is the tendency to create ``singularity clusters'' and non-smooth solutions. These two artifacts result from mixing the scale with the phase, dampening large rotations by reducing magnitudes. To counter that, recent methods worked with the Ginzburg-Landau functional~\cite{Viertel:2019, palmer:2024}, a smoothness measure on unit-length fields. Factoring out magnitude makes fields rotationally smooth and consequently avoids singularity clusters. However, Ginzburg-Landau methods have not been demonstrated yet on higher-order representations such as FEEC~\cite{arnold:2006} or piecewise-polynomial~\cite{Stein:2020, boksebeld_high-order_2022}. This is not a trivial generalization since normalizing the associated nodes does not lead to a unit-length field in the interpolants. Polar methods such as ours optimize for phase differences directly, resulting in rotational smoothness by design.

\paragraph{Polar methods} In comparison, polar methods have been studied much less. Earlier methods~\cite{Bommes:2009, Li:2006} represented fields with direct phases. However, this required working with integers encoding the period jumps. Other methods sidestepped this by working with \emph{differential} phase~\cite{Ray:2008, Crane:2010}, encoding the rotations between neighboring frames, and thus controlling vector field curvature (singularities) explicitly, often at the cost of a single linear system. These methods have been presented for low-order interpolants in either vertex-based (e.g.,~\cite{Palacios:2007}), or face-based (e.g.~\cite{Crane:2010}) settings, restricting the positioning of singularities. Our method is also a differential phase method, where we generalize to a piecewise power-linear setting, allowing for the placement of singularities anywhere on a mesh. Polar methods have been used for intuitive temporal interpolation between fields~\cite{chen:2023, Solomon:2019}, and as a logarithmic representation of conformal maps~\cite{weber:2010}, relating scale and rotation through the Hilbert transform. One interesting class of methods is using rotation angles to guide moving frames~\cite{Corman:2024, Coiffier:2023}, which has been extended to volumes~\cite{Corman:2019}. We note two hybrid works (mixing polar and Cartesian) that have connections to our method: In~\cite{Knoppel:2015}, stripe patterns are computed with a Cartesian representation, and then texture coordinates are interpolated inside triangles by mimicking a holomorphic function, reproducing high frequencies. A similar representation has been taken by~\cite{chen:2023}, which represents both frequency (by phase differentials) and the actual phase as the argument of a complex number (both correlated by least squares). However, neither controls singularity placement directly nor offers a higher-order interpolant. Liu \textit{et al.}~\shortcite{Liu:2016} encoded corner and edge-based phases with a continuous interpolant on the entire mesh like us. However, they only worked with vertex-based fields and have not demonstrated the ability to prescribe singularities everywhere. Mitra \emph{et al.}~\shortcite{Mitra:2024} also uses a (primal)-edge-based phase (but not scale), and only deals with singularities of index $\pm 1$, for generating valid knitting structures. They target the global consistency of the resulting foliations. This is an interesting application that might be possible to combine with our representation.
\paragraph{Higher-order fields} To the best of our knowledge, higher-order fields have only been presented in the Cartesian setting. The focus is on designing function spaces that interface with finite-element scalar spaces in a structure-preserving manner, either as piecewise polynomials~\cite{boksebeld_high-order_2022, Manded:2020}, or subdivision fields~\cite{deGoes:2016sec, Custers:2020}. Our representation cannot use polynomial spaces for phases, as they are not well-defined around singularities. Instead, we control the polar components of a power-linear field. 

%% file: sections/piecewise-linear-polar-fields.tex
\section{Piecewise Linear Polar Fields}
\label{sec:piecewise-linear-polar-fields}

\subsection{Background: Piecewise-constant polar fields}
\label{subsec:trivial-connections}
We describe a common variant of differential polar fields, trivial connections~\cite{Crane:2010}. Consider a $2$-manifold triangle mesh $\M = \left\{\V, \E, \F\right\}$ with an arbitrary genus and any number of boundary loops. We denote the discrete Gaussian curvature (angle defect) as $\kappa:|\mathcal{V}|\times 1$ and assign an arbitrary, but fixed, basis in each triangle face $f \in \F$, defining a local complex plane. This defines a \emph{connection dual $1$-form}, which is an angle-based quantity $r_{fg}$, encoding the rotation of the base of face $f$ to that of an adjacent face $g$. Oriented sums of (dual)-edge quantities around each vertex are encoded by the $0$-form differential matrix $d_0$~\cite{Desbrun:2005} where $d_0^Tr = \kappa$ is the \emph{holonomy} of the vertex. In the case of genus and boundaries, we augment $d_0^T$ to accommodate the independent cycles (Sec.~\ref{subsec:boundaries-and-genus}).
We consider fields of a single vector per face, parameterized as complex numbers $u$ in the respective face bases. Two vectors $u_f$ and $u_g$ are \emph{parallel} when $u_g = e^{i\cdot r_{fg}}u_f$. The consequence of holonomy is that fields cannot be perfectly parallel around cycles. We obtain as-parallel-as-possible fields, by finding rotation angles $\theta:|\mathcal{E}|\times 1$ such that $u_g = e^{i\cdot (r_{fg}+\theta_{fg})}u_f$. $\theta$ ``cancels'' the holonomy up to full rotations of $2\pi I$, where $I:|\mathcal{V}|\times 1$ are the \emph{indices} of the cycles. The resulting system is:
$$
\theta = \argmin |\theta|^2\ \  s.t.\ \ d_0^T\theta = 2\pi I - \kappa.
$$
Trivial Connections is simple and guarantees that singularities result exactly where prescribed and nowhere else; however, it is limited by placing singularities only at vertices. Furthermore, the piecewise-constant field is poorly aliased for high-index singularities, and non-smooth for bad meshes (Figs.~\ref{fig:teaser} and~\ref{fig:comparison-trivial-connection}). We next show how to mitigate these effects using higher-order fields.

\subsection{Representation}
\label{subsect:representation}
We introduce piecewise-linear single-vector fields on simply-connected meshes, and generalize them to non-simply-connected meshes in Sec.~\ref{subsec:boundaries-and-genus} and power fields in Sec.~\ref{subsec:power-linear fields}. We denote pointwise quantities in bold (e.g.,  $\vectheta(z)$), and integrated quantities in regular  (e.g., $\theta_e$).

\begin{figure}
\includegraphics[width=0.3\textwidth]{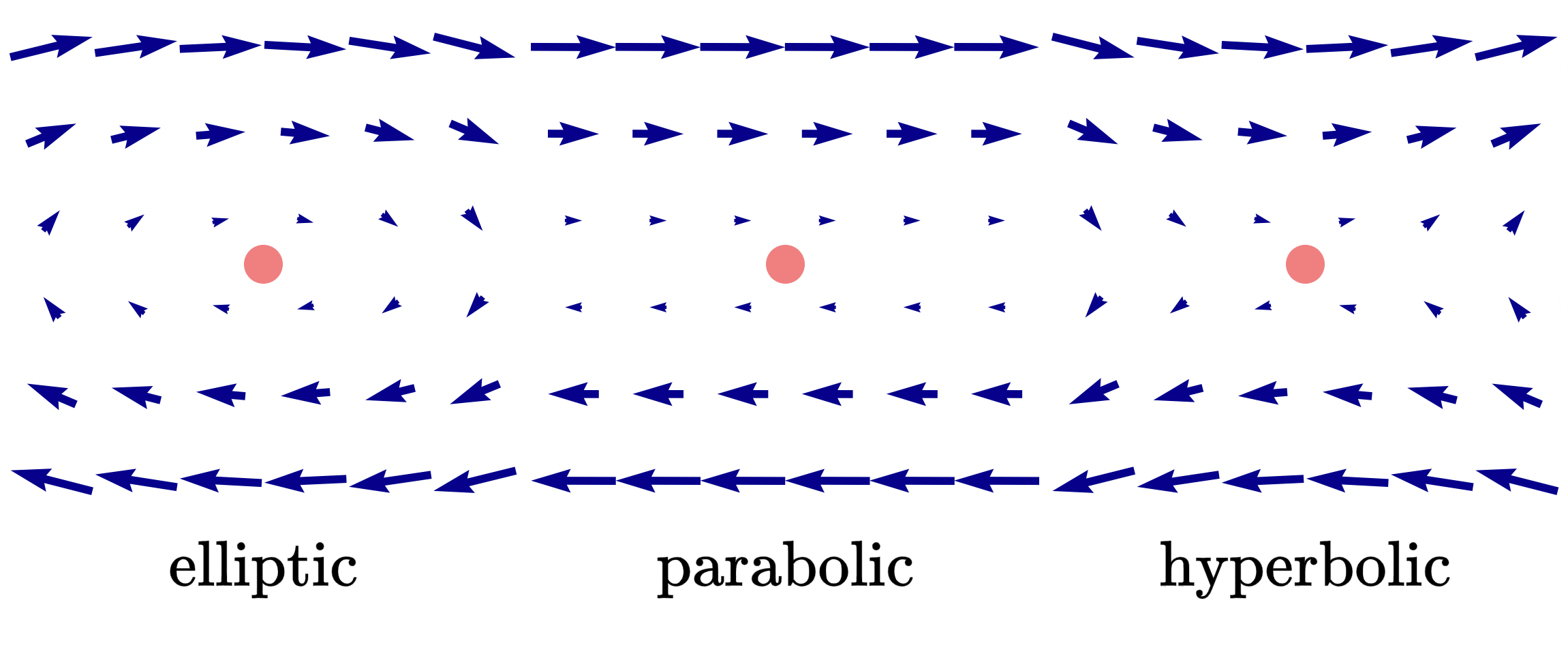}
\caption{(Left) An elliptic field with a $+1$ singularity, (middle) a linear singularity, and (right) a parabolic field with a $-1$ singularity.}
\label{fig:linear-field-types}
\end{figure}

\paragraph{Piecewise-linear fields} We define a \emph{piecewise linear field} as a field that comprises piecewise face-based vector fields as:
\begin{align*}
\forall f=ijk \in \F,\ \forall z \in f,\ \ \mathbf{u}(z) &= a_fz+b_f\overline{z}+c_f\\ 
&= B_i(z)u_i+B_j(z)u_j+B_k(z)u_k,
\end{align*}
where $z \in \mathbb{C}$ is in the local complex coordinates,  $a_f, b_f, c_f \in \mathbb{C}$ are the coefficients of the complex linear function,  $B_{i|j|k}(z)$ are barycentric coordinates, and $u_i, u_j$ and $u_k$ are the degrees of freedom on the face corners. These fields are not generally $C^0$-continuous across edges. This representation corresponds to a complex formulation of the space $\mathcal{X}_1$ in \cite{boksebeld_high-order_2022}. The field is \emph{holomorphic} when $b_f=0$, and \emph{anti-holomorphic} when $a_f=0$\footnote{We note a similar characterization was provided in~\cite{Liu:2016}.}.
Linear fields have a rich geometry, inherited from their quadratic field lines. A singular point (or \emph{singularity}) $s$ is where $\mathbf{u}(s) = 0$. A linear field can have one singular point, no singular points (when $a_f=b_f=0$), or a full singular line (when $|b_f| = |a_f|$). The singularity may fall within the triangle, in which case $f$ is singular, or otherwise \emph{regular}. Using the conic section definitions, we denote the field as \emph{elliptic} when $|a_f| > |b_f|$, and the singularity has index $1$ (see auxiliary material for a proof), \emph{hyperbolic} when $|b_f| > |a_f|$ and the singularity has index $-1$, and \emph{parabolic} when $|b_f| = |a_f|$ (Fig.~\ref{fig:linear-field-types}). We avoid the parabolic case in this paper, although it can be used for design.  The ratio of $|a_f|$ to $|b_f|$ determines the \emph{eccentricity} of the field, measuring how "stretched" or "compressed" a conic section is relative to a perfect circle or hyperbola. It is known~\cite{Djerbetian:2016, Stein:2020} that the Whitney forms, interpolating edge-based $1$-forms linearly, only reproduce perfect holomorphic ($b_f=0$) fields.

\paragraph{Phase gradient and scale} The field is only continuous on faces, where discontinuities happen across edges and vertices. Discontinuities are necessary for non-trivial parallel transport. There is, regardless, a consistent definition of edge and vertex cycle curvature with explicit jump encoding. We are interested in explicitly encoding the phase of the field, to control singularities. We consider a local environment $\Omega \subset f$ for some $f \in \mathcal{F}$, which does not contain a singular point. The complex logarithm is well-defined in $\Omega$:
\begin{align}\label{eq:log_linear}
    \forall z \in \Omega,\ \bld{\varsigma}(z) + i\bld{\psi}(z) = \log(\bld{u}(z)).
\end{align}
$\log(\bld{u}) = \log|\bld{u}|+i\ \text{atan2}(Im(z), Re(z))$, where $\bld{\varsigma}$ is the logarithm of the scale, and $\bld{\psi}$ is the phase of the field. Phase is a branched function in the presence of singularities. Thus, polar representations (e.g.,~\cite{Ray:2006, Ray:2008, Crane:2010}) usually consider a \emph{differential} representation: for any patch that does not include a singularity, we can ``comb'' $\bld{\psi}$ to a continuous function, and define $\vectheta = \nabla \bld{\psi}$. $\vectheta$ is invariant to the combing, and is well-defined everywhere away from singularities. For a linear field: 
$$
\vectheta(z,\overline{z}) = Im\left(\frac{\nabla \bld{u}(z)}{\bld{u}(z)}\right) = \frac{Im\left((a_f+ b_f)\overline{\bld{u}(z)}\right)}{|\bld{u}(z)|^2}
$$
$\vectheta$ has a pole at the singularity, where the phase is undefined. Inversely, $\vectheta$ defines a linear field up to scale and global phase; that means it has 4 degrees of freedom (against the full 6 of linear fields).

\begin{figure}
\centering
\includegraphics[width=0.3\textwidth]{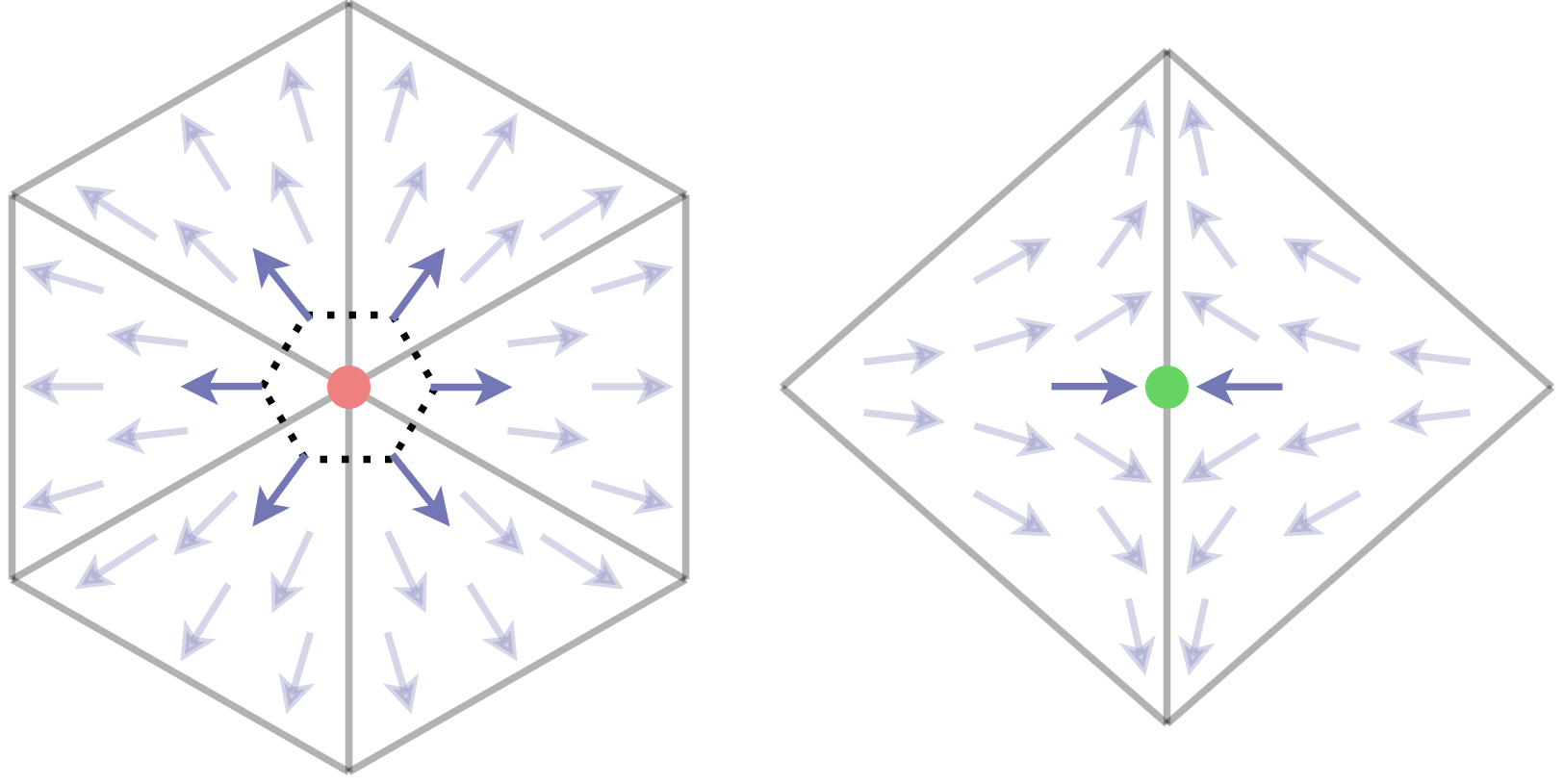}
\caption{Vertex (index $I_v = 1$) and edge ($I_e = -1$) singularities.}
\label{fig:vertex-edge-singularities}
\end{figure}

\paragraph{Vertex and edge singularities} Consider the boundary of a region $\partial \Omega$, oriented counterclockwise. We have that:
\begin{equation}
    \label{eq:index}
\oint_{\partial \Omega}\vectheta 
  \cdot d(\partial\Omega) = 2\pi I_\Omega - \kappa_\Omega,
\end{equation}
where $I_\Omega$ is the \emph{index} of the field in $\Omega$. This integral definition allows for piecewise continuity, making it possible to define singularities on edges and vertices. Given an edge $e=ki$ adjacent to left face $f_1=ijk$ and right face $f_2=kli$, and suppose both faces have the same coordinate system, we parameterize the edge as $e(t) = (1-t)z_k+tz_i$, and define the jump $\theta_e(t)$ such that away from singularities $\theta_e(t) = \text{arg} \frac{\bld{u}_2(e(t))}{\bld{u}_1(e(t))}$. This is invariant to the choice of phase branch. The jump can either be inferred (using principal matching~\cite{boksebeld_high-order_2022}) or represented explicitly, as in all differential polar methods. Vertex singularities (Fig.~\ref{fig:vertex-edge-singularities} left) are defined in terms of the jumps around the corners of a vertex, reproducing trivial connections (Sec~\ref{subsec:trivial-connections}). Edge singularities of index $I_e$ are defined where $\theta_e(t) = \pi I_e,\ I_e \in \mathbb{Z}$, resulting in codirectional (opposite for odd $I_e$) vectors on both sides (Fig.~\ref{fig:vertex-edge-singularities} right). Piecewise continuity means that the scale doesn't have to be zero at the singularity.

\subsection{Phase and scale forms}

\begin{figure}
\includegraphics[width=0.35\textwidth]{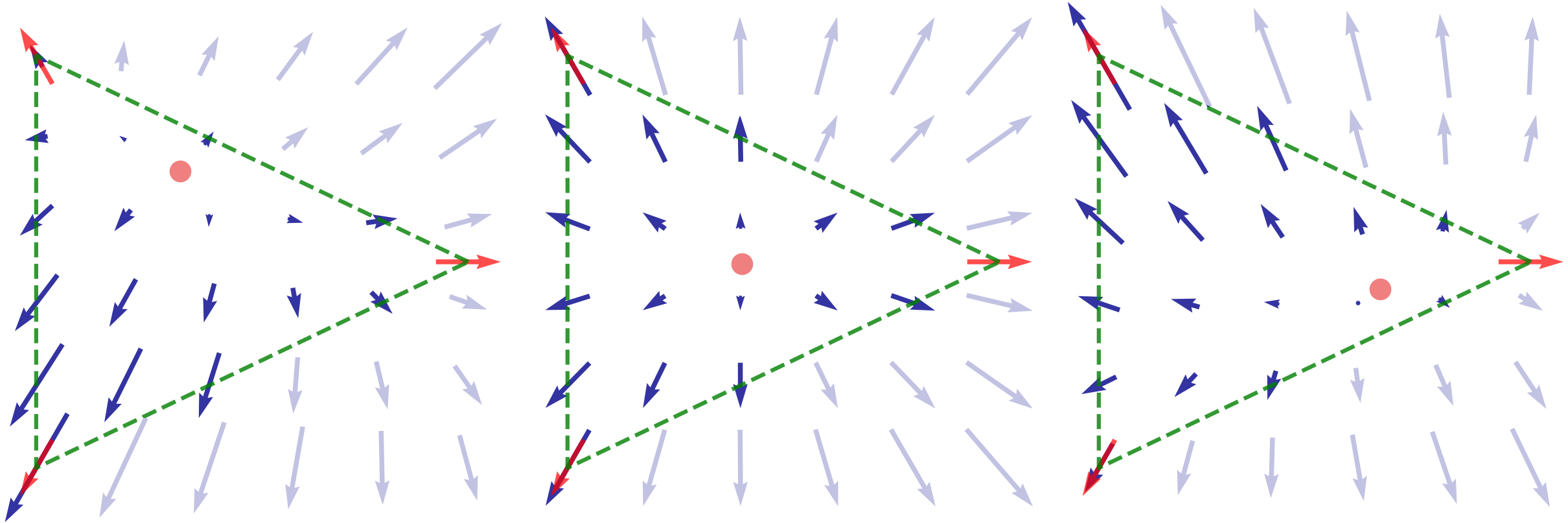}
\caption{There is a $2$-dimensional space of fields with the same edge $\theta$, differing in the location of the singularity (red). We show three such examples.}
\label{fig:same-theta-different-field}
\end{figure}

\paragraph{The phase $1$-form} To control the phase gradient effectively, we encode it as a discrete $1$-form on triangle edges. For a triangle $f= 
 ijk$, we define the moment $\theta_{ij}$ of $\vectheta$ on edge $e=ij$ as:
$$
\theta_{ij} = \int_{e}{\vectheta \cdot de}
$$
The integral in Eq.~\ref{eq:index} is then discretized into $\theta_{ij}+\theta_{jk}+\theta_{ki} = I_f$, the index of the triangle. For a linear field, this is the index of the only singularity that might be within the triangle ($I_f=0$ otherwise). 
The integrated edge-based $\theta$ representation is comfortable and concise as it allows to prescribe triangle indices easily. However, the transformation $\vectheta  \rightarrow \theta$ is \emph{lossy}, as $\theta$ only has $2$ degrees of freedom (per choice of $I_f$); this means there is a $2$-dimensional space of fields for each $\theta$. These manifest as either choosing the location of the singularity, or determining the eccentricity of the field (Fig.~\ref{fig:same-theta-different-field}).
\paragraph{The scale $0$-form} Working with singularity location or eccentricity is unintuitive, especially for regular faces. Instead, an elegant way to provide these degrees of freedom without limiting the generality of the representation is to add edge-based scale differentials. One might consider using edge-based $1$-forms of the log-scale differential $\varsigma_{e} = \int_e\bld{\varsigma}\cdot de$, which must be consequently \emph{closed} (summing to zero in the face), complementing the missing $2$ degrees of freedom. However, This is not an ideal representation; singularities within faces make $\varsigma_e$ unbounded from below, and consequently not a numerically stable representation. We instead use the original scale $\vecsigma = |\bld{u}| = \exp(\bld{\varsigma})$  differentials. This completes the missing $2$ dofs. The disadvantage is that scale is not invariant to addition by constant, which biases the optimization; however, as we explain in Sec.~\ref{subsec:optimization}, this only changes the scale smoothness and not the adherence to the singularity or alignment constraints. Scale is an exact form (unlike $\theta$), and thus we encode it as a $0$-form with corner-based values $\sigma_{i,j,k}$ in $f$.

Given the scale $0$-form and the phase $1$-form, and an assignment of phase $\psi_i$, without loss of generality, we reproduce the field as:
\begin{align}
\nonumber u_i &= \sigma_i \exp(i\psi_i),\\
    \nonumber u_j &= \sigma_j \exp(i(\psi_i+\theta_{ij})),\\
    u_k &=\sigma_k\exp(i(\psi_i+\theta_{ij}+\theta_{jk})).
    \label{eq:interpolant}
\end{align}
These definitions are for a single triangle; both $\theta$ and $\sigma$ are generally discontinuous across edges and are by definition discontinuous around vertices with non-trivial Gaussian curvature. We note that the use of scale here is to parametrize the missing degrees of freedom; one can normalize the field pointwise, to obtain a unit-length field adhering to the prescribed phase. We next introduce a combinatorial structure that homogenizes the definition of singular cycles, simplifying our representation and optimization.

%% file: sections/bevelled-meshes.tex
\section{Fields on Bevelled Meshes}
\label{sec:fields-on-bevelled-meshes}
The dual $1$-form representation of trivial connections does not generalize directly; there are more degrees of freedom of jumps across edges, and singularities could appear anywhere. In the following, we devise an alternative representation that captures these properties.
\begin{figure}
\centering
\includegraphics[width=.42\linewidth]{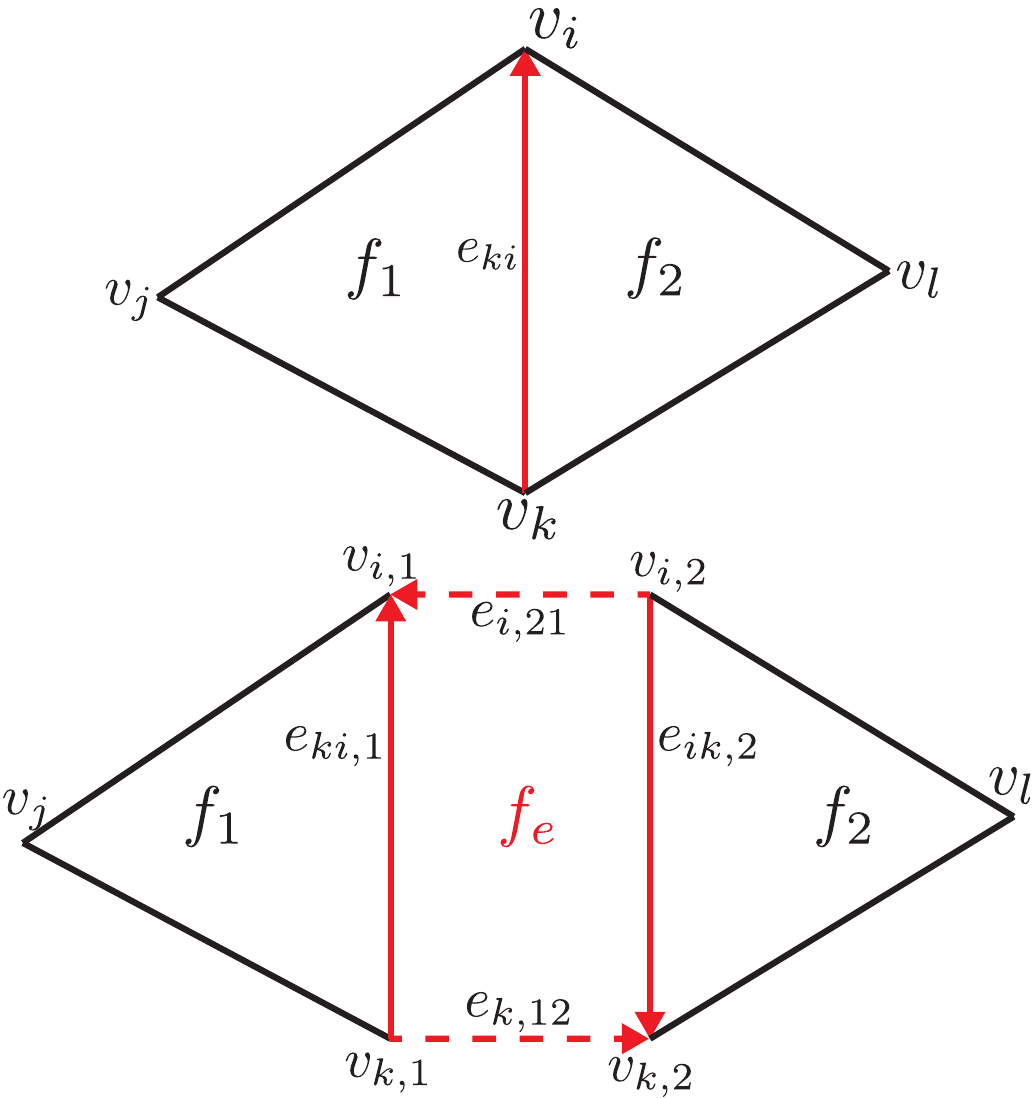}
\hspace{0.3In}
\includegraphics[width=.28\linewidth]{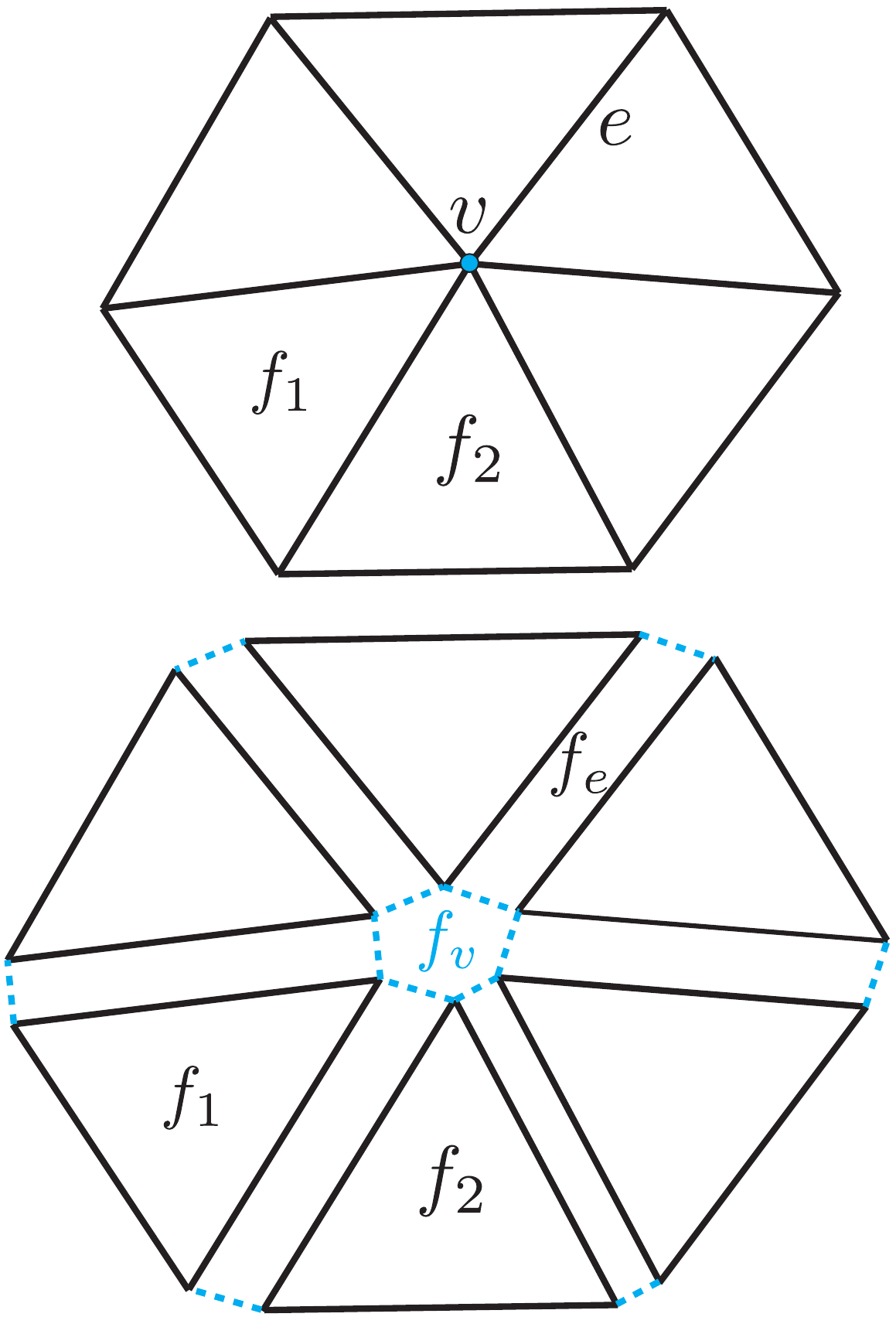}
\caption{Beveling edges (left) and vertices (right)}
\label{fig:bevelled-meshes}
\end{figure}
\paragraph{Beveled meshes}

We \emph{bevel} $\mathcal{M}$ into $\overline{\mathcal M} = \{\overline{\V}, \overline{\E}, \overline{\F}\}$, by creating a virtual edge-face $f_e \in \overline{\mathcal{F}}$ for every edge $e \in \mathcal{E}$ and vertex-face $f_v \in \overline{\mathcal{F}}$ for every vertex $v \in \mathcal{V}$ as in Fig.~\ref{fig:bevelled-meshes}. There are two new edges in $\overline{\mathcal{E}}$ for every non-boundary edge in $\mathcal{E}$, and two new edges between every two original adjacent corners. We denote the former as \emph{split} edges and the latter as \emph{jump} edges. 
We then have $\overline{\mathcal{F}} = \overline{\mathcal{F}}_\mathcal{V} \bigcup \overline{\mathcal{F}}_\mathcal{E} \bigcup \mathcal{F}$.
These new elements are purely combinatorial and do not change the geometry of the mesh. The advantage of the beveled mesh $\overline{\mathcal{M}}$ is that \emph{all} singularities are encoded as face cycles $f \in \overline{\mathcal{F}}$, and where $\sigma$ is a $\overline{\mathcal{V}}$-valued $0$-form and $\theta$ is a $\overline{\mathcal{E}}$-valued $1$-form. $\theta$ values on the jump edges encode the discontinuities between adjacent corners in the original $\mathcal{M}$. Linear fields are only defined on original $\mathcal{F}$.

\subsection{Fields from Index Prescription}
\label{subsec:fields-from-index-prescription}
\paragraph{User-guided design}
A user prescribes a set of singular points $s \in S$ anywhere on $\mathcal{M}$, with a choice of index $I(s \in S)$. The set of indices must conform to the index theorem: $\sum_{s \in S}{I_s} = \sum_{v\in \mathcal{V}}{\kappa}  = \chi_{\mathcal{M}}$, the Euler characteristic. There is also a limit of one singularity per mesh element, where face singularities must have index $\pm 1$ (this limit is removed for power fields in Sec.~\ref{subsec:power-linear fields}). 
\paragraph{Cycle Constraints} We encode local cycles on $\overline{\mathcal{M}}$ with the  $d_1:|\overline{\mathcal{F}}| \times |\overline{\mathcal{E}}|$ operator, defined as follows:
$$
d_1(f,e) = \begin{cases}
    \pm1 & \text{$e \in \overline{\mathcal{E}}$ positively/negatively coincident to $f \in \overline{\mathcal{F}}$}, \\
    0 & \text{otherwise},
\end{cases}
$$
We extend the Gaussian curvature to $\overline{\kappa}$ on the beveled faces $\overline{\mathcal{F}}$, where $\overline{\kappa}_{f_v} = \kappa_v$, and trivially $\overline{\kappa}_{f\in \mathcal{F}} = 0$ and $\overline{\kappa}_{f \in \overline{\mathcal{F}}_e} = 0$. Given the per-face index vector $I:|\overline{\mathcal{F}}|\times 1$, the cycle constraint on $\theta$ is:
\begin{equation}
    d_1 \theta = 2\pi I - \overline{\kappa}.
\label{eq:theta-cycle-consistency}
\end{equation}

\paragraph{Reproducing singularities}
Recall that $\theta$ does not encode a unique linear field, and thus we need extra priors on the field. We compute desired values for $\theta$, and resulting constraints for $\theta$ and the scale $\sigma$, that reproduce singularities. For face singularities, we assume that the field should be \emph{as isotropic as possible}, meaning either as \emph{holomorphic} (for $I_f = 1$, a perfect circle) or as \emph{anti-holomorphic} (for $I_f = -1$, a perfect hyperbola) as possible.  Consider the set of singular faces ${\mathcal{S}} \subset \overline{\mathcal{F}}$, and from which the subset of singular original faces $\mathcal{S}_\mathcal{F} \subset \mathcal{S}$. We collect the $\theta^f$ values that result from assuming isotropy (see Auxiliary material for the calculation), and incorporate them in our optimization (Sec.~\ref{subsec:optimization}). We use superscripts to denote desired isotropic values for $\theta$ and constrained values for $\sigma$. Given computed $\theta$ values in a singular original face, the $\sigma$ values are entirely determined (up to a face-wise multiplicative factor) by the position of the singularity. Denote these values as $\sigma^{i,j,k}$ for face $f=ijk$, our constraint is then (similarly for $jk$ and $ki$):
\begin{equation}
\sigma_i\sigma^j - \sigma_j\sigma^i = 0,
\label{eq:scale-face-constraint}
\end{equation}
\begin{wrapfigure}{r}{0pt}
    \includegraphics[width=.4\linewidth]{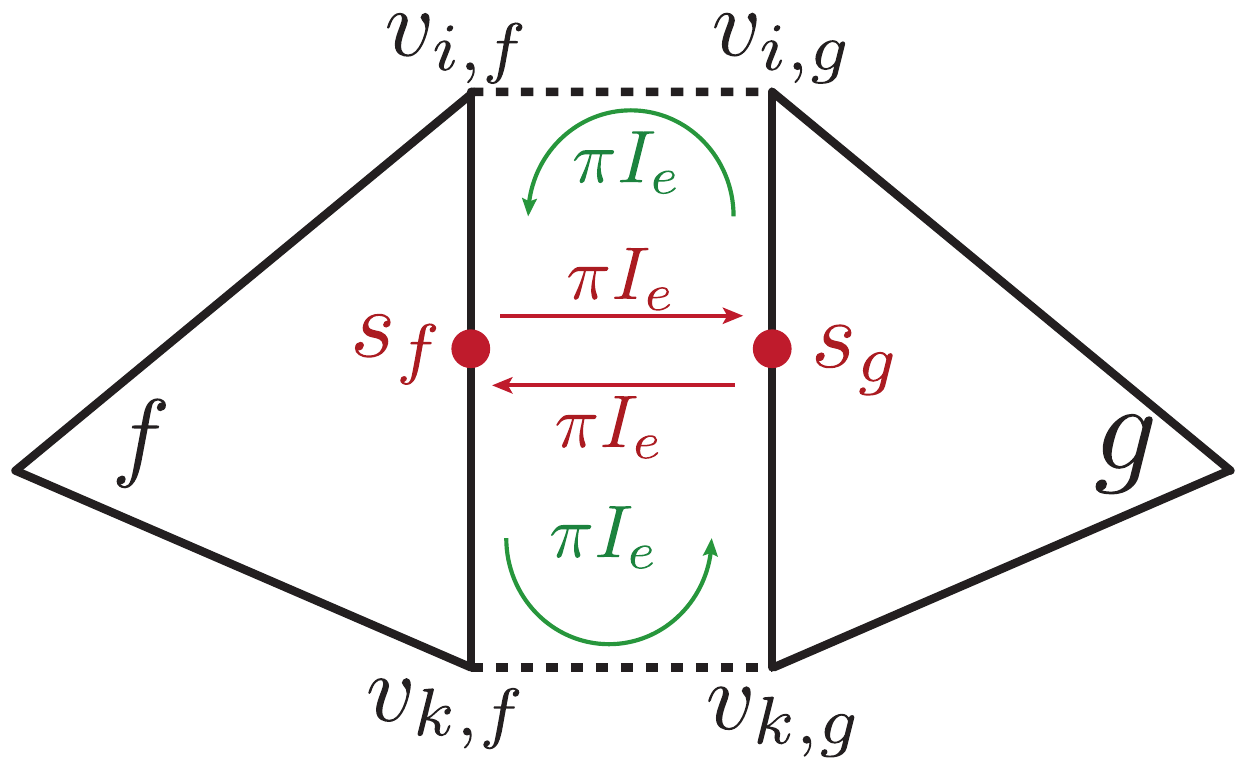}
\end{wrapfigure}
For edge singularities, beveled to singular edge-faces $\mathcal{S}_\mathcal{E} \subset  {\mathcal{S}}$, we need to reproduce the singularity at the prescribed location on the edge, by enforcing the $\pm\pi I_e$ jump of the field across the edge-face. We constrain the jump at the singular point $s$ by breaking the edge-face $f_e$ into two half-cycles (see inset), where we need to produce two $\theta$ values for each broken split edge, to obtain:

\begin{equation}
\theta_{sk,f}+\theta_{k,fg}+\theta_{ks,g} = \theta_{si,g}+\theta_{i,gf}+\theta_{is,f} = \pi I_e.
\label{eq:part-edge-consistency}
\end{equation}
Rather than using explicit extra variables, we compute the part-edge $\theta_{sk,f}, \theta_{is,f}$ (resp. for $g$) as a post-process. We cannot add any more scale variables (as the field is linear in the faces). Thus, to reproduce these part-edge $\theta$ values, we must impose constraints on the scale as for face singularities. If $s_f = (1-t)v_{k,f}+tv_{i,f}$ (resp. $g$) we get:
\begin{align}
\nonumber  &\sigma_{k,f}(1-t)\sin(\theta_{sk,f})+\sigma_{i,f}\cdot t\cdot \sin(\theta_{si,f}) = 0 \Rightarrow \\
&\sigma^{k,f} = -t\sin(\theta_{si,f}),\ \sigma^{i,f} = (1-t)\sin(\theta_{sk,f})
\label{eq:scale-edge-constraint}
\end{align}
and similarly for $g$. We then add constraints for every singular edge in the form of Eq.~\ref{eq:scale-face-constraint}. Aside from the constraints, we also have a prior for edge singularities, so that they are as \emph{intrinsically} isotropic as possible on the flap, and for which we collect the isotropic $\theta^e = \theta^f \bigcup \theta^g$ values. There are no scale constraints for vertex singularities, which comprise only jump edges. We however also collect the isotropic values for $\theta^v$, assuming the vertex is intrinsically isotropic. See auxiliary material for further details.


\subsection{Optimization} \label{subsec:optimization}
\paragraph{Objectives} Aside from adhering to the singular-cycle constraints (Eq.~\ref{eq:theta-cycle-consistency}) 
, and consequent scale constraints (Eq.~\ref{eq:scale-face-constraint} including Eq.~\ref{eq:scale-edge-constraint}), we want the field to be 1) as smooth as possible everywhere, 2) as isotropic as possible in singular elements.   We collect all scale constraints in matrix form: $C_\theta\sigma = 0$, and all the isotropy values collected from singular faces and edges to a vector $\theta^\mathcal{S}$.
Our optimization system is then:
\begin{align}
\nonumber (\theta, \sigma) &= \text{argmin}\int_{\overline{\mathcal{M}}}{\left(|\vectheta|^2+|\nabla \vecsigma|^2\right)} + \lambda_{\mathcal{S}} \int_{\mathcal{S}}{\left| \vectheta- \vectheta^\mathcal{S}\right|^2},\\
\nonumber \text{s.t.}\ \  d_1\theta &= 2\pi I - \nonumber  \overline{\kappa}, \\
C_\theta \sigma &= 0,\ \ \ \sigma \geq \epsilon.
\label{eq:continuous-optimization}
\end{align}
The last inequality guarantees that all resulting scales are positive; we use $\epsilon = 10^{-6}$ and $\lambda_\mathcal{S} = 50$.

\paragraph{Discretization} Since we have jump discontinuities, we choose a finite-volume approach to discretizing the norms in Sys.~\ref{eq:continuous-optimization}. The primary challenge is that $\vectheta$ and $\vecsigma$ are coupled in a nonlinear way. We relax the norm by approximating $\vecsigma$ as piecewise linear, and $\vectheta$ as piecewise constant, as if they are linear Whitney forms~\cite{Desbrun:2005}. We note this only affects the metric governing the objective function, but not the linear constraints or the post-process interpolation. We consider a general piecewise-linear real scalar quantity $\bld{\alpha}$, and a flap with faces  $f = ijk, g = kli$ and edge $e=ki$. Applying Stokes' theorem, we get:

\begin{equation}\label{eq:finite-volume}
    \int_{f \bigcup g}{\nabla \bld{\alpha}} = \int_{ f}{\nabla 
 \bld{\alpha}} + \int_{g}{\nabla  \bld{\alpha}} + \int_e{(\bld{\alpha}_f - \bld{\alpha}_g)\cdot \hat{e}^\perp}, 
\end{equation}
$\bld{\alpha}_f$ and $\bld{\alpha}_g$ are the values on $e$ on both sides, and $\hat{e}^{\perp}$ is the edge $e$ rotated by $\frac{\pi}{2}$ and normalized. When $\bld{\alpha}$ is piecewise linear, and its gradient is piecewise constant, we get:
$$
\int_{f \bigcup g}{\nabla \bld{\alpha}} =A(f)\nabla \bld{\alpha}(f) + A(g)\nabla \bld{\alpha}(g) + \frac{1}{2}\left(\alpha_{l,i}-\alpha_{r,i}+\alpha_{l,k}-\alpha_{r,k}\right)e_{ki}^{\perp}.
$$
$A(f)$ is the area of face $f$ (resp. for $g$). Since $\nabla \bld{\alpha}$ is differential, it is possible to express it purely in terms of edge and jump differentials $\alpha_{ij} = \alpha_j-\alpha_i$ and $\alpha_{i,rl} = \alpha_{i,l}-\alpha_{i,r}$, as:
\begin{align*}
\nabla \bld{\alpha}(f) &= \frac{e_{ij}^\perp\alpha_k+e_{jk}^\perp\alpha_i+e_{ki}^\perp\alpha_j}{2A(f)}\\
&=\frac{e_{ki}^\perp - e_{jk}^\perp}{6A(f)}\alpha_{ij}+\frac{e_{ij}^\perp - e_{ki}^\perp}{6A(f)}\alpha_{jk}+\frac{e_{jk}^\perp - e_{ij}^\perp}{6A(f)}\alpha_{ki}. \\
\end{align*}
We construct a matrix $D:|2\mathcal{E}|\times|\overline{\mathcal{E}}|$ so that $D\cdot d_0\alpha$ is the (vector-valued) integral for every flap (explicit construction is in the auxiliary material. The Dirichlet energy of $\bld{\alpha}$ is:
$$
\int_\mathcal{M}|\nabla \bld{\alpha}|^2 \approx |Dd_0\alpha|^2 = \alpha^T \left(d0^TD^T M_\mathcal{E}Dd_0\right)\alpha^T = \alpha^TL\alpha.
$$
$M_\mathcal{E}$ is a diagonal matrix of (vector) \emph{inverse} flap weights (as $D$ gives an integrated value):
$$
M_\mathcal{E}(2e,2e) = M_\mathcal{E}(2e+1,2e+1) = \frac{1}{A(f)+A(g)}.
$$
In our setup, $\bld{\sigma}$ behaves like $\bld{\alpha}$ and $\theta$ behaves like $d_0\alpha$.
Empirically, we noticed that this favors more constant gradients and bigger jumps, which is likely due to the metric approximation. To mitigate this, and drawing from discontinuous Galerkin methods~\cite{arnold:2000}, we penalize the jump by a coefficient $\lambda_J \geq 1$:
$$
\int_{f \bigcup g}{\nabla \bld{\alpha}} = \int_{ f}{\nabla 
 \bld{\alpha}} + \int_{g}{\nabla  \bld{\alpha}} + \lambda_J\int_e{(\bld{\alpha}_f - \bld{\alpha}_g)\cdot \hat{e}^\perp}, 
$$
We use $\lambda_J = 50$. The advantage of this approximation is that it allows us to decouple the optimization into $\theta$ and $\sigma$ steps.
For matching the isotropic values in the singular elements, we use a variant of $D$ and $M$. We create $D_\mathcal{E} = D[\mathcal{E},:]$ (slicing rows), $D_\mathcal{V} = D[e \in N(\mathcal{V}),:]$ (all adjacent flaps to each vertex) and $D_\mathcal{F}:|\mathcal{S}_\mathcal{F}|\times |\overline{\mathcal{E}}|$ which is just the face contributions $A(f)\nabla \bld{\alpha}$ for each singular (original) face. Then we construct $D_\mathcal{S}=[D_\mathcal{V};D_\mathcal{E};D_\mathcal{F}]$. $M_\mathcal{S}$ has the respective inverse flap areas for $D_\mathcal{V}$ and $D_\mathcal{E}$ and inverse face areas for $D_\mathcal{F}$.
\paragraph{Optimizing for $\theta$} Our system for $\theta$ is a quadratic minimization for smoothness and isotropy with the cycle constraints as follows:
\begin{align}
\nonumber \theta &= \text{argmin}\left[\theta^TD^TM_\mathcal{E}D\theta + \lambda_\mathcal{S}(\theta_\mathcal{S}-\theta^\mathcal{S})^TD_\mathcal{S}^TM_\mathcal{S}D_\mathcal{S}(\theta_\mathcal{S}-\theta^\mathcal{S})\right],\\ 
d_1\theta &= 2\pi I - \overline{\kappa}.
\label{eq:theta-optimization}
\end{align}
This is solved by a single linear system, with LU decomposition.

\paragraph{Optimizing $\sigma$} We solve for:
\begin{align}
\nonumber \sigma &= \text{argmin}\left[\sigma^Td_0^TD^T M_\mathcal{E} Dd_0\sigma\right],\\
\text{s.t.}\ \ C_\theta \sigma &= 0,\ \ 
\sigma \geq \epsilon.
\label{eq:sigma-optimization}
\end{align}
This is solved as a single convex system using CVX.

\begin{figure*}
\centering
\includegraphics[width=0.8\textwidth]{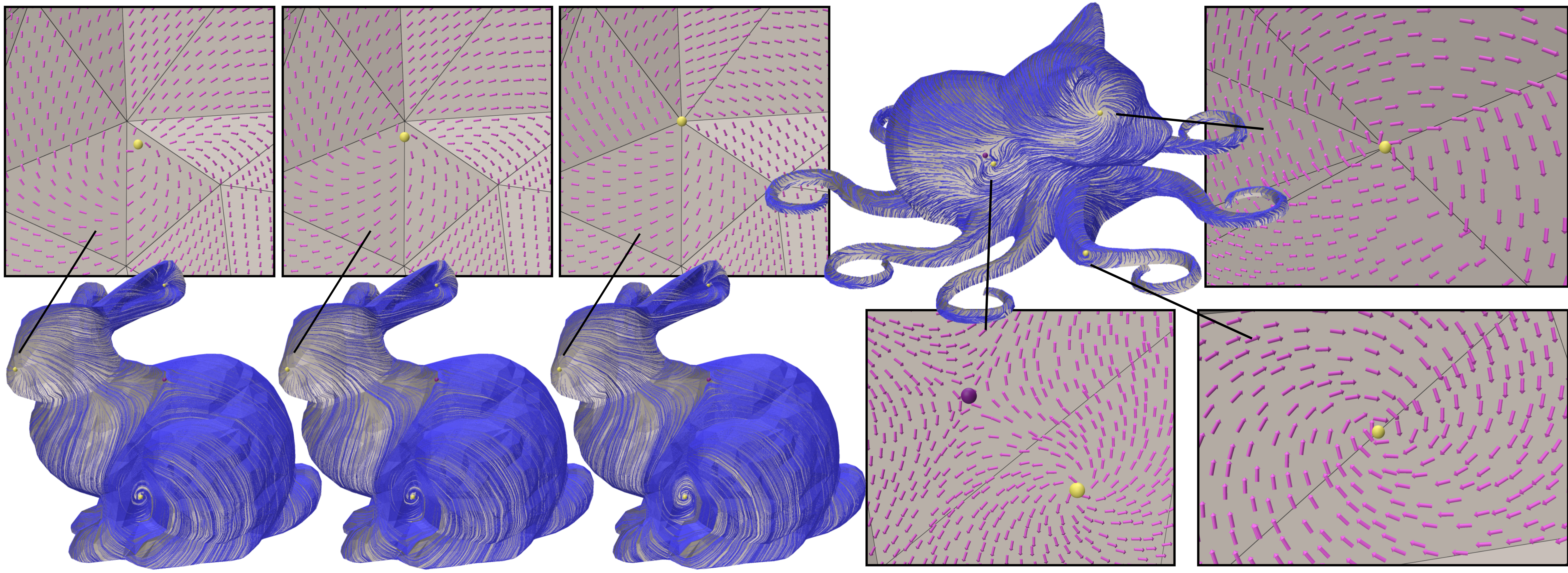}
\caption{Our linear field on simply-connected meshes. Our result is stable to small perturbations (left), and results in smooth fields for different indices (right).}
\label{fig:simply-connected-results}
\end{figure*}

\paragraph{Global Integration} Having computed $\theta$ and $\sigma$, we need to integrate the entire field by determining corner values $u$ at the original faces. This could be done by flood-filling from a single chosen corner. It's however more stable to solve the following linear system:
$$
\forall e=ij \in \overline{\mathcal{E}},\ \ \sigma_i u_j - \sigma_j u_i\cdot e^{i(\theta_{ij}+r_{ij})} = 0,
$$
where $r_{ij}$ is the connection form (Sec.~\ref{subsec:trivial-connections}). We show some results for linear fields on simply connected meshes in Fig.~\ref{fig:simply-connected-results}.

\subsection{Boundaries and Genus}
\label{subsec:boundaries-and-genus}
\begin{figure}
\centering
\includegraphics[width=0.3\textwidth]{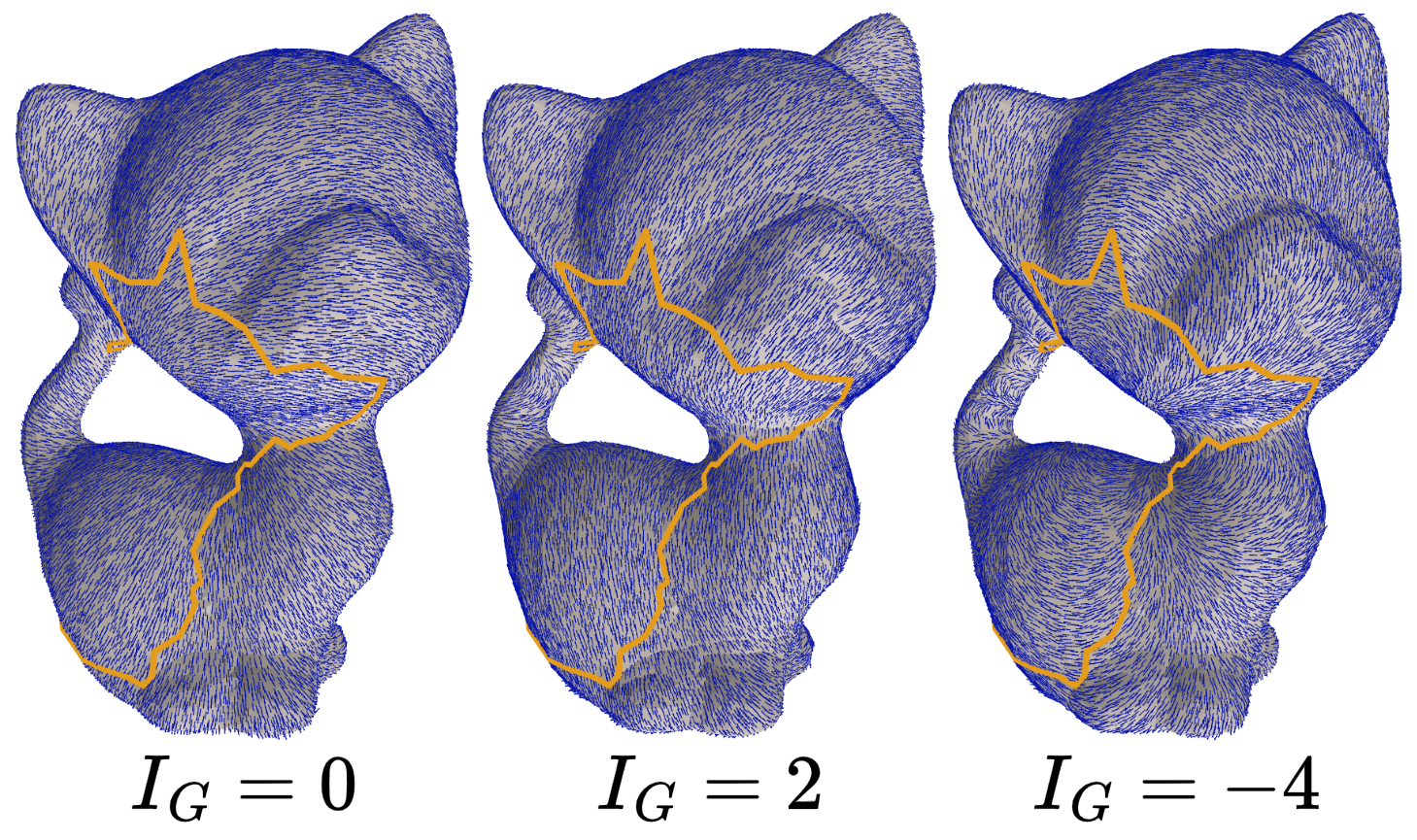}
\caption{A higher-genus example, where we vary the index $I_G$ of a cycle.}
\label{fig:high-genus}
\end{figure}

\begin{figure}
\centering
\includegraphics[width=0.4\textwidth]{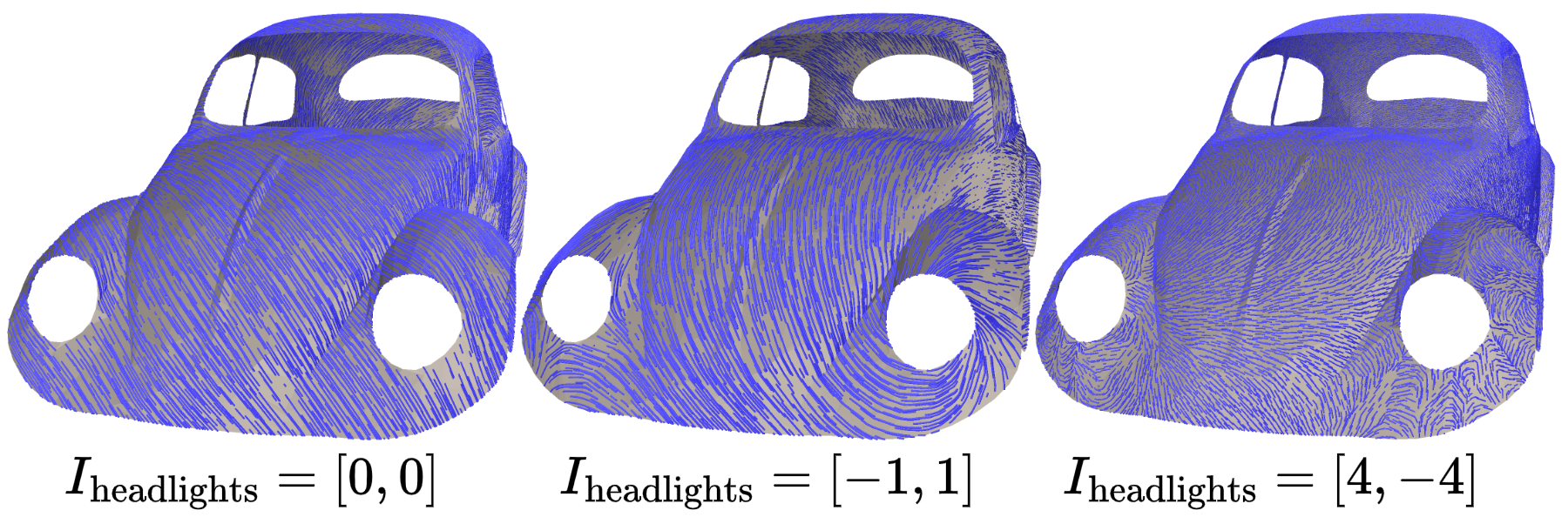}
\caption{A mesh with boundaries, where we vary the indices of the two front boundaries inversely to each other.}
\label{fig:boundaries}
\end{figure}

\paragraph{Higher-genus meshes}
In the presence of genus-based (co)homology cycles, we need to adapt Sys.~\ref{eq:theta-optimization}. We consider the set $\mathcal{H}$ of homological cycles on $\overline{\mathcal{M}}$, where each $h \in \mathcal{H}$ is an independent simple non-contractible loop of edges, and where $|\mathcal{H}| = 2g = 2-\chi(\mathcal{M})$. We construct the cohomology operator $H: |\mathcal{H}|\times |\overline{\mathcal{E}}|$ from oriented edges to homological cycles, similarly to $d_1$:
$$
H(h,e) = \begin{cases}
    \pm1 & \text{$e \in \overline{\mathcal{E}}$ positively/negatively coincident to cycle $h$}, \\
    0 & \text{otherwise},
\end{cases}
$$
The curvature $\kappa_h$ of a cycle $h$ is the one-sided angle defect, measured as $\sum_{v\in \mathcal{H}}{\left[\pi - \sum{\alpha}_v\right]}$ for all angles $\alpha_v$ on all (original) vertices along one side of the path. Given a prescription of indices $I_\mathcal{H}$, we add $H\theta = 2\pi I_\mathcal{H} - \kappa_\mathcal{H}$ to the $\theta$ optimization Sys.~\ref{eq:theta-optimization}. $I_\mathcal{H}$ are independent and do not count towards the index sum. We show results in Fig.~\ref{fig:high-genus}.
\paragraph{Boundaries} Boundaries are similar to genus, where we define an operator $B:|\mathcal{B}|\times |\overline{\mathcal{E}}|$ that encodes the signed boundary loops $\mathcal{B}$ and add them to Sys.~\ref{eq:theta-optimization}. There are two differences: 1) boundaries count toward the index sum, as 
$\chi(\mathcal{M}) = |\mathcal{V}|-|\mathcal{E}|+|\mathcal{F}| = 2-2g-|\mathcal{B}|.$, and 2) the rows corresponding to boundary vertex- and edge-faces should be removed from $d_1$. We show some examples in Fig.~\ref{fig:boundaries}.

\subsection{Power-linear fields}
\label{subsec:power-linear fields}
\paragraph{Higher indices}
The indices of edge and vertex singularities can be arbitrary since we have jump edges. However, face singularities for linear fields can only admit $\pm 1$ indices. To enable higher-order singularities in faces, we consider \emph{power-linear} fields of the form:
$$
\bld{u}(z) = \left(a_fz+b_f\overline{z}+c_f\right)^{I_f}, \forall f \in \mathcal{F},\ I_f \in \mathbb{Z}\setminus 0,
$$
where $I_f$ is the prescribed singularity index. We exclude $0$ since constant fields are a subset of linear fields. The power preserves the singularity location, but ``folds'' the field on itself $I_f$ times; as a consequence, $\vectheta$ (and its edge moments $\theta$) are scaled by $I_f$, and the scale $\vecsigma$ is taken to the power of $I_f$ relative to the original linear field. Our algorithm mostly proceeds like before to optimize for $\theta$ and $\sigma$, when the isotropic values are scaled by $I_f$. One difference is in the post-process interpolant, modifying Eq.~\ref{eq:interpolant}:
\begin{align*}
u_i &= \sigma_i^{\nicefrac{1}{I_f}} \exp(i\psi_i),\\
    u_j &= \sigma_j^{\nicefrac{1}{I_f}} \exp(i(\psi_i+\nicefrac{\theta_{ij}}{I_f})),\\
    u_k &=\sigma_k^{\nicefrac{1}{I_f}} \exp(i(\psi_i+\nicefrac{\left(\theta_{ij}+\theta_{jk}\right)}{I_f})),\\
    \bld{u}(p) &= \left(B_i(p)u_i+B_j(p)u_j+B_k(p)u_k\right)^{I_f},
\end{align*}
The effect is that of reproducing the underlying linear field, and taking the pointwise power of $I_f$. A more major difference is that we need to compute $I_f$ for non-singular faces. For this, we interpolate $I_f$ smoothly on $M$ after computing $\theta$. Aside from the known singularity faces, for each face $f \in \mathcal{F}$ adjacent to a singularity of index $I$ (whether it's an edge or a vertex), we fix this as its $I_f$. We further fix $I_f$ for regular faces where any $|\theta|>I_f\pi$, since they cannot have a lower index by design. We interpolate $I_f$ to all non-fixed faces $f \in \mathcal{F}$ by minimizing the Dirichlet energy with the $L_2 = d_1M_Id_1^T$ Laplacian, with the dual weights of~\cite{Brandt:2018}:
$$
M_I(e,e) = \frac{3l_e^2}{A(f)+A(g)}.
$$
$l_e$ is the edge length of $e$. We then round $I_f$ to the nearest integer. We show the results of higher-order mixed indices in Fig.~\ref{fig:high-index}.

\paragraph{N-symmetric fields} We often want to represent an $N$-field (in the categorization of~\cite{Vaxman:2016}), where in every point we have $N$ symmetric vectors. The most common are $N=3,4,6$, used to construct triangle, quad, and hexagonal meshes, respectively. $N$-fields have \emph{fractional} singularities of index $\frac{I}{N}$, for $I \in \mathbb{Z}$. We represent them by \emph{power fields}, with a complex vector $\bld{U}(z)$ for every point $z$, such that its set of $N^{th}$ roots is the $N$-vector at $z \in \mathbb{C}$:
$$
\bld{U}(z) = \bld{u}(z)^N \Rightarrow  \bld{u}(z) = \left\{\sqrt[N]{\bld{U}(z)}\cdot \exp\left(\frac{2\pi i}{N}\right) \Big \vert \  \forall i \in \mathbb{Z}\right\}
$$

Power fields reduce the representation of $N$-fields to that of the \emph{power vectors} $\bld{U}$. In the polar setup, $N$-fields don't induce any change to the algorithm: given a prescription of singularities of indices $\frac{I}{N}$, we proceed like higher-order singularities to obtain fields that are interpolated as $\bld{u}(z) = \left(a_tz+b_t\overline{z}+c_t\right)^{\nicefrac{I_f}{N}}$. Note that prescribed singularities should sum up to the topological invariant with the fraction. The only difference is that $N$ is the same for the entire mesh. We show results of $N$-fields in Fig.~\ref{fig:power-fields}.




%% file: sections/results.tex
\section{Experiments}
\label{sec:experiments}
\paragraph{Implementation details} Our method is implemented in Python with \texttt{scipy} for sparse linear algebra. The streamlines are generated using \texttt{Directional}~\cite{Directional} and visualized with \texttt{Polyscope}~\cite{polyscope}. We worked on a 2019 Macbook with an Intel Core $i5$ processor and $8$ GB of memory. For a 38k-face mesh, it takes approximately $1.1$ minutes to compute the field, where about 45 seconds are spent precomputing.

\paragraph{Comparison to trivial connections} We compare our algorithm to Trivial Connections by~\cite{Crane:2010} in Fig.~\ref{fig:comparison-trivial-connection}, where we put our singularities only on vertices, for fair comparison. The singularities are of a variety of indices. For a quantitative comparison, we evaluate the phase Dirichlet energy $E = \theta^T L \theta$ (Sec.~\ref{subsec:optimization}), averaged by the number of faces, and where we upsample it uniformly to its three corners. 
 Piecewise-constant vectors evidently cannot sustain the large rotations needed by the singularities, and this results in visually apparent artifacts, whereas ours reproduces these rotations nicely, even in the presence of uneven triangulations. 

\paragraph{Mesh quality} Our representation results in similar fields when applied to different mesh qualities in Fig.~\ref{fig:mesh-quality}. We attribute this to the higher-order representation with explicit smooth rotations.

\begin{figure}
\centering
\includegraphics[width=.8\linewidth]{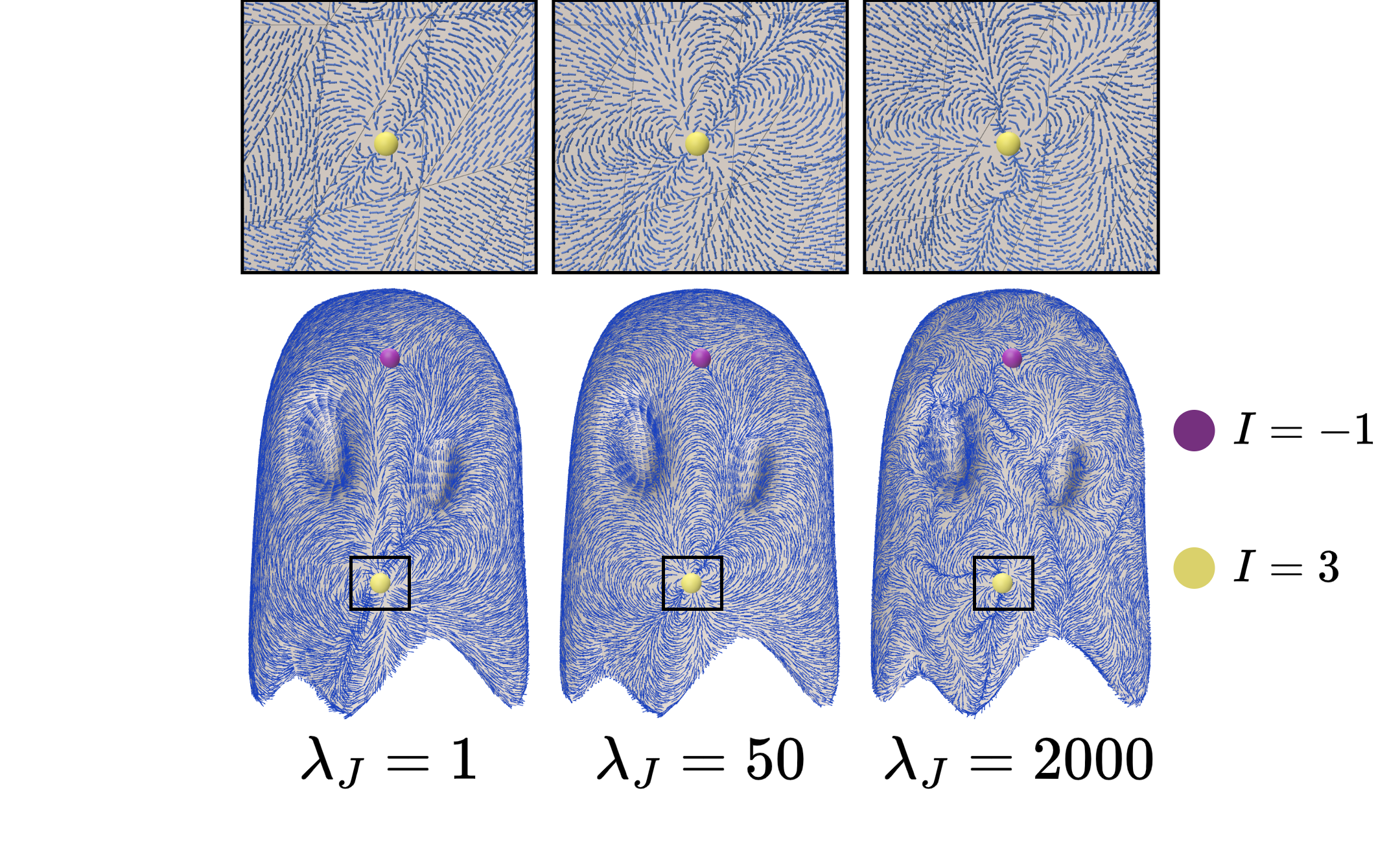}
\caption{Ablating the jump penalty $\lambda_J$.}
\label{fig:jump_ablation}
\end{figure}

\begin{figure}
\centering
\includegraphics[width=.95\linewidth]{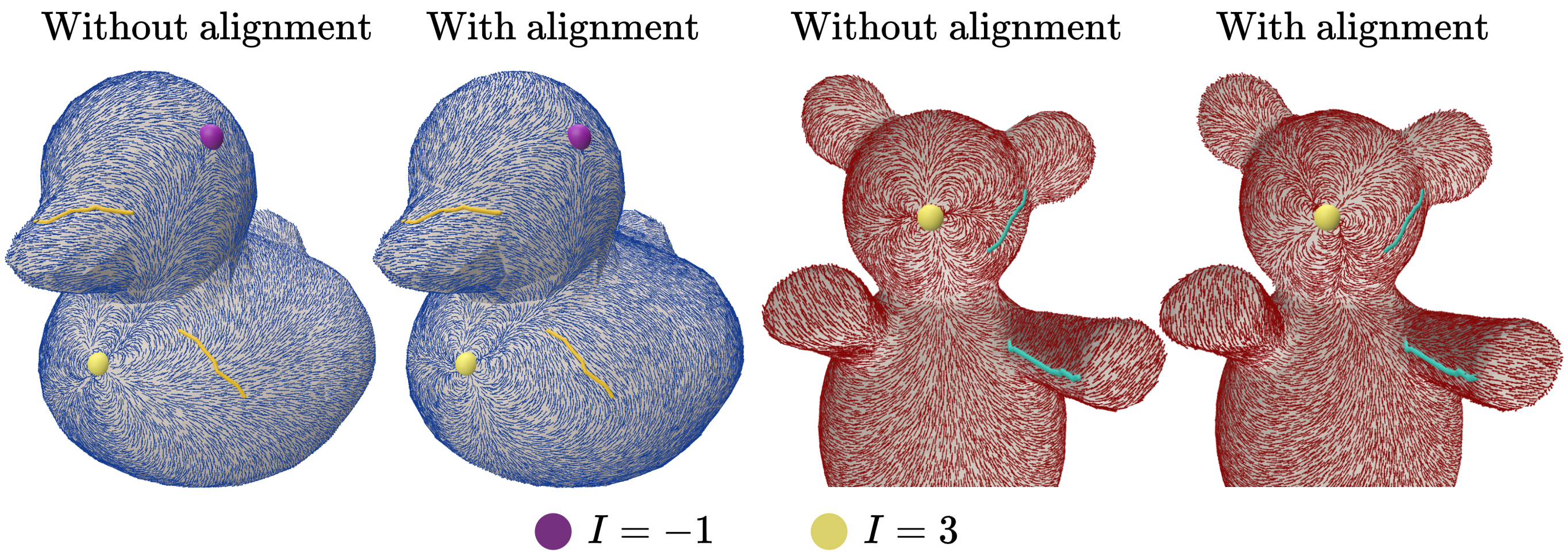}
\caption{Aligning the field to prescribed curves.}
\label{fig:alignment}
\end{figure}

\paragraph{Ablating $\lambda_J$} We test different jump penalties $\lambda_J$, as demonstrated in Fig.~\ref{fig:jump_ablation}. A low value results in smooth faces with large jumps, and a very big value suppresses the necessary jumps at the expense of face smoothness. As such, our choice is reasonable.
\paragraph{Alignment} We can align the field to features along curves (Fig.~\ref{fig:alignment}). This is done by a combination of forcing halfcycles on faces where an alignment curve passes, as for edge singularities, and adding the differential of a global vertex-based field, to synchronize between the curves. We give exact details in the auxiliary material.

%% file: sections/conclusion.tex
\section{Discussion}\label{sec:discussion}
\paragraph{Higher-order fields} Our work focused on power-linear fields.  However, Cartesian fields like~\cite{boksebeld_high-order_2022} include quadratic and higher-order fields. These fields are problematic in the polar setting; to exemplify this, consider the following quadratic field, for some $\delta$: $\bld{u}(z) = (z-\delta)(\overline{z}+\delta),\ \  \lim_{\delta \rightarrow 0 }{\bld{u}(z) = |z|^2}.$
The field has a $+1$ singularity at $\delta$ and a $-1$ singularity at $-\delta$. Any triangle enveloping the singularities would seem regular and smooth. This means we cannot guarantee the exact reproduction of indices. Linear fields are a big improvement over constant fields, with higher orders having diminishing returns in approximation quality.
\paragraph{Limitations} As with other polar methods, it is difficult to encode differential quantities such as curl or divergence. We \emph{require} the user to explicitly input singularities, and do not offer a way to devise them automatically by energies such as the Ginzburg-Landau functional. It would be interesting to explore a hybrid version of our representation alongside a Cartesian one in future work.

%% file: appendix.tex






\section*{Appendix}
In the following, we provide extended proofs and implementation details, referenced in the main document.

\section{Field Properties}
\subsection{Singularity indices} 
A linear field with a singularity at $s \in \mathbb{C}$ can be written as:
$$
\bld{u}(z) = a(z-s) + b(\overline{z}-\overline{s}) = az+b\overline{z}+c,
$$
where $c = -sa-\overline{s}b$. In the simple case of a linear field, the jacobian (expressed in real $\mathbb{R}$ coordinates) is constant on the plane, where: 
$$
J_{2\times 2} = \begin{pmatrix}Re(a+b) & -Im(a+b) \\ Im(a-b) & Re(a-b)\end{pmatrix}
$$

The index of the singularity depends on $\operatorname{sign}(\det(J))$, where the field is parabolic iff $\det(J)=0$. We have:
$$
\det(J) = |a|^2-|b|^2.
$$
\subsection{Calculating isotropic $\theta^f$} 
For a face $f=ijk$ with a singularity of index $I_f$ at $s \in \mathbb{C}$, the ideal holomorphic power-linear field is, $\bld{u}(z) = (z-s)^{I_f}$ for $I_f>0$. The ideal anti-holomorphic field is $\bld{u}(z) = (\overline{z}-\overline{s})^{I_f}$, for $I_f<0$. In both cases, the rotation $\theta^f_{ij}$ on the edge is the classic winding-number formula:
\begin{equation}
    \theta^{f}_{ij} = I_f\cdot\angle jsi = I_f \cdot \text{arg}\left(\frac{\left(z_s-z_i)(\overline{z_s-z_j}\right)}{|z_s-z_i|\cdot|z_s-z_j|}\right),
    \label{eq:face-to-edge-moments}
\end{equation}
which is computed for $\theta_{jk}^f$ and $\theta_{ki}^f$ similarly. 
For edge singularities on edge $e=ki$ adjacent to faces $f=ijk$ and $g=kli$, we simplify collect the individual face coefficients $\theta^f$ and $\theta^g$ of both faces, including zero jump variables $\theta_{i,fg}=\theta_{k,fg}=0$. This describes isotropic fields that are intrinsic on the flap. This also means that one must prescribe $\theta_{f,ik} = \theta_{g,ki} = \pi I_f$ as the correct branch resolution of the respective argument functions.

\begin{figure}
    \includegraphics[width=0.4\textwidth]{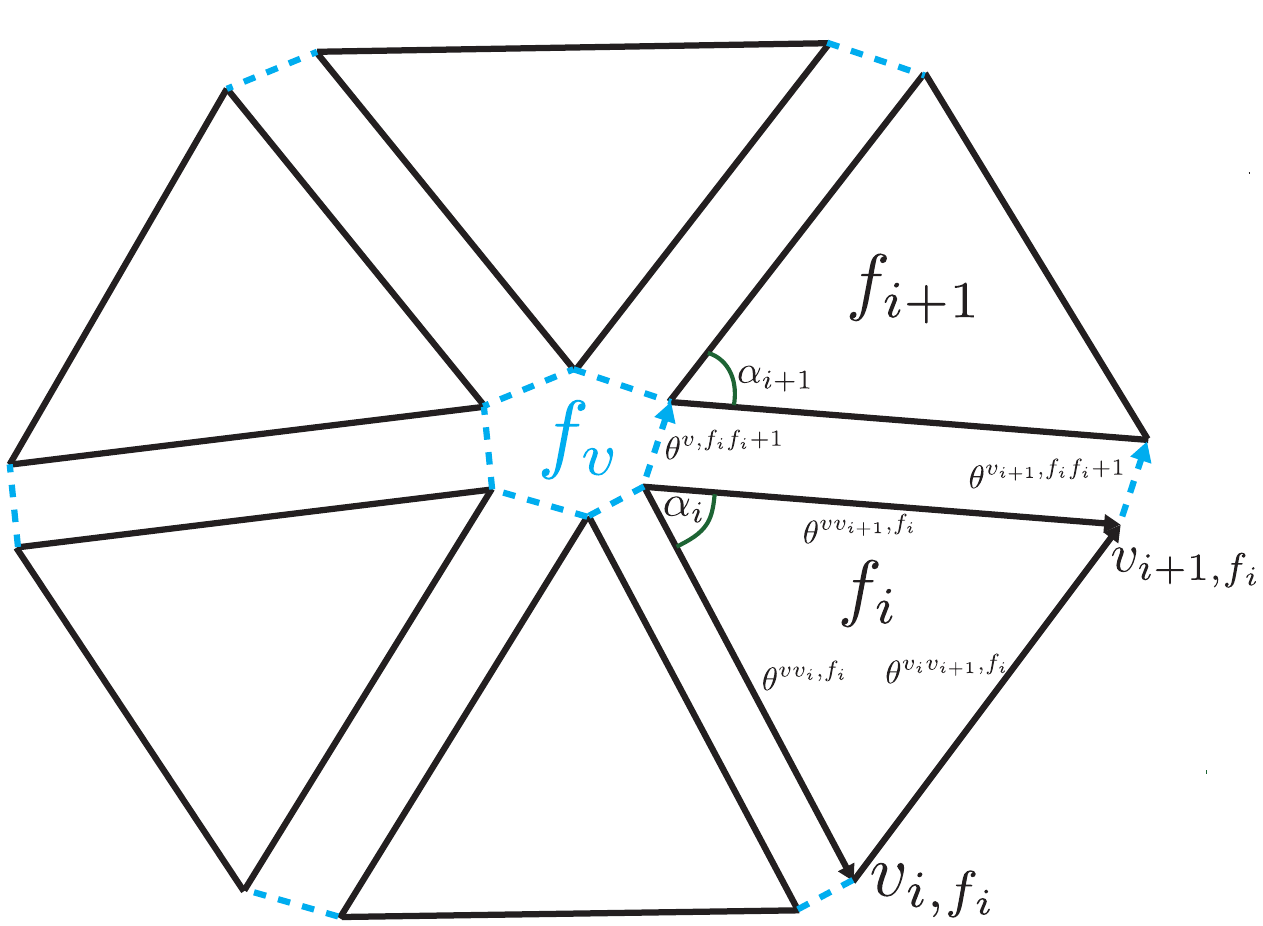}
    \caption{Notation for vertex singularity $\theta$ isotropy prescription.}
    \label{fig:vertex-singularities}
\end{figure}

\paragraph{Vertex singularities} For a vertex singularity at some $v \in 
\mathcal{V}$, we would like to generate as much as possible a continuous isotropic field of the index $I_v$. One must have non-zero jumps at the vertex itself, since $\theta$ is not continuous at the singularity and the field is just zero at the point. Furthermore, the vertex has curvature, which means there is no ``'ideal'' solution intrinsically. We prescribe the following for $\theta^v$, which is a reasonable solution where $\kappa_v = 0$ to approximately reproduce $(z-v)^{I_v}$ (see Fig.~\ref{fig:vertex-singularities}):
\begin{enumerate}
    \item For any outer edge $e_i = (v_i,v_{i+1})$, we set $\theta^{v_iv_i+1,f_i}$ as the winding-number (Eq.~\ref{eq:face-to-edge-moments}), scaled by the total rotation around $f_v$, which is $2\pi I_v - \kappa_v$.
    \item For any two split outgoing edges $(v, v_{i,f_i})$ and $(v, v_{i+1,f_i})$ adjacent at face $f_i$, we set the same $\theta$. Since the index adds up to $0$ in each face, we have:
    $$
    \theta^{vv_i,f_i} = -\theta^{vv_{i+1},f_i} = -\frac{1}{2}\theta^{v_iv_{i+1},f_i}.
    $$
    \item The jump variables at the outer edges are set to zero: \mbox{$\theta^{v_{i+1}, f_if_{i+1}} = 0$}
    \item We divide the total rotation among the singularity corner jump variables, weighted according to flap angles:
    $$
    \theta^{v,f_if_{i+1}} = \frac{\alpha_i+\alpha_{i+1}}{\sum_i{\alpha_i+\alpha_{i+1}}}\left(2\pi I_v - \kappa_v\right)
    $$
    
\end{enumerate}

\subsection{Scales from phases}
Given a face singularity at position $s$, and computed $\theta_{ij,jk,kl}$ values (not necessarily isotropic), we need to compute $\sigma^{i,j,k}$ values that would reproduce $s$, and that are collected into the singularity constraints (Eq. 5 in the original paper). Assume an arbitrary phase $\psi_i$, to which the corner field values are:
\begin{align*}
u_i &= \sigma^ie^{i\psi_i},\\
u_j &= \sigma^j e^{i(\psi_i+\theta_{ij})},\\
u_k &= \sigma^k e^{i(\psi_i+\theta_{ij}+\theta_{jk})},
\end{align*}
and where $s = B_iz_i+B_jz_j+B_kz_k$, for barycentric coordinates $B_{i,j,k}$. We then have the linear relation
$$
\bld{u}(s) = B_iu_i+B_ju_j+B_ku_k = 0.
$$
This is a system with two variables (one complex) for the three $\sigma$ variables. As scale is defined up to a multiplicative factor, we just take any vector from its kernel as $\sigma^{i,j,k}$.

\subsubsection{Edge singularities} 
\label{subsubsec:edge-singularities}
As explained in the main paper (Sec. 3.2), edge singularities do not have to have zero scale due to the jump discontinuities. There are nevertheless conditions on the scales at the corners of the edge for reproducing the opposite phases on both sides.

\begin{figure}
    \includegraphics[width=0.4\textwidth]{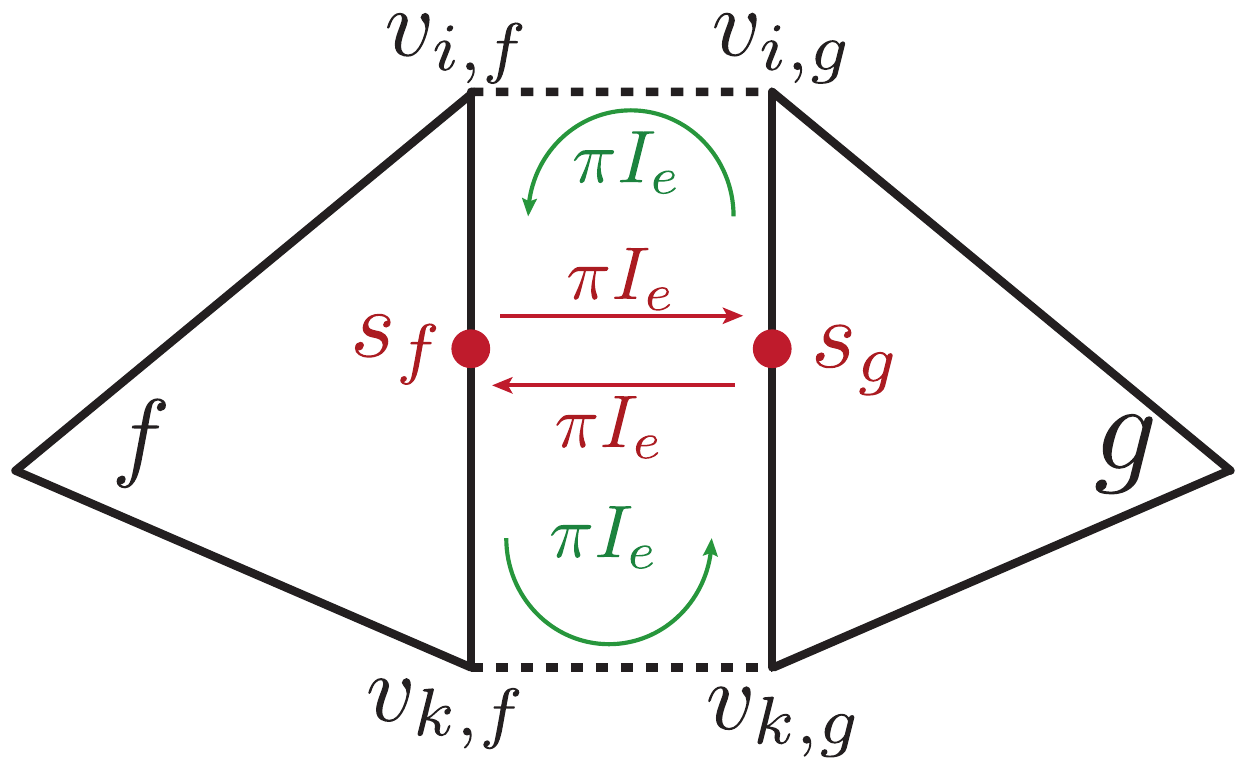}
    \caption{Notation for edge singularity $\theta$ isotropy prescription.}
    \label{fig:edge-singularities}
\end{figure}

\paragraph{Computing part-edge $\theta$} Assume a singularity of index $I_e$ on edge $e_{ij}$ parameterized by $s_f = (1-t)v_{k,f}+tv_{i,f}$ (resp. $s_g$) as in Fig.~\ref{fig:edge-singularities}, and where $\theta_{ik,f},\theta_{k,fg}, \theta_{ki,g}, \theta_{i,gf}$ have been computed for all edges, respecting the edge singularity cycles and as perfect as possible. We next compute the part-edge values. 

Recall that we need to compute part-edge quantities so that: 
\begin{equation}
\theta_{sk,f}+\theta_{k,fg}+\theta_{ks,g} = \theta_{si,g}+\theta_{i,gf}+\theta_{is,f} = \pi I_e.
\label{eq:part-edge-consistency}
\end{equation}

Consider edge-face adjacent to faces $f, g$, and an assignment of phases $\psi$ for the corners.

\begin{wrapfigure}{r}{0pt}
    \includegraphics[width=.25\linewidth]{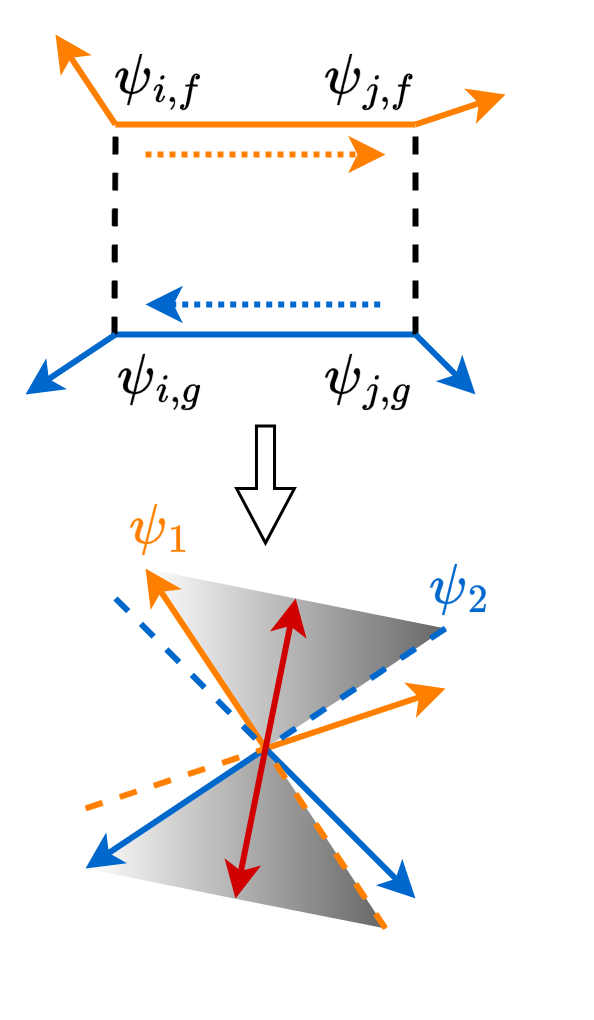}
\end{wrapfigure}

Without loss of generality, we locally integrate the corner phases $\psi_{i,f},\psi_{k,f}, \psi_{k,g}, \psi_{i,g}$ (see inset).  We begin by deciding on a set of intermediate phases $\{\psi_{s_f}, \psi_{s_g}\}$ such that $\psi_{s_f} =  \psi_{s_g} + I_e\pi$ on the unfolded flap. We want the interpolated field to align with these phases at $s_e$ so that $s_e$ behaves as a singularity. Since $\psi_{s_f}$ must lie within $(\psi_{i, f}, \psi_{j,f})$ and $\psi_{s_g} = \psi_{s_f}+I_e\pi$ be within $(\psi_{j, g}, \psi_{i,g})$ (all phases are defined under the same local basis on the unfolded flap), to make sure the choice of $\psi_{s_f}$ leads to a feasible $\psi_{s_g}$, the available range of $\psi_{s_f}$ would be the narrowest one of $(\psi_{i, f}, \psi_{j,f}), (\psi_{i, f}, \psi_{i, g} + I_e\pi), (\psi_{j, g} + I_e\pi, \psi_{j,f}), (\psi_{j, g} + I_e\pi, \psi_{i, g} + I_e\pi)$ (see inset for $I_e = 1$ case). With $(\psi_1, \psi_2)$ denoting this range, we then select the middle phase:
$$
\psi_{s_f} = \frac{\psi_1 + \psi_2}{2}
$$
as this would be the orientation at $s_e$ if the field is perfectly isotropic. Consequently $\psi_{s_g} = \psi_{s_f}+I_e\pi$. 
\paragraph{Scale constraints}
The part-edge $\theta$ values are not used as explicit variables in our system, and thus scale is the only thing that can constrain the full edge $\theta$ to divide in the prescribed manner. In face $f$, the constraints on scale can be written as follows in terms of achieving the interpolated phase $\psi$:

\begin{align*}
e^{i\psi_{s_f}} &= \sigma^{k,f}(1-t)e^{i\psi_{k,f}} + \sigma^{i,f}te^{i\psi_{i,f}}\Rightarrow\\
1 &= \sigma^{k,f}(1-t)e^{i\theta_{sk,f}} + \sigma^{i,f}te^{i\theta_{si,f}}\\
1 &= \sigma^{k,g}(1-t)e^{i\theta_{sk,g}} + \sigma^{i,g}te^{i\theta_{si,g}}.
\end{align*}
In fact, since the global scale prescription at $e$ is arbitrary, the minimal necessary and sufficient form of these constraints is:
\begin{align*}
Im\left(\sigma^{k,f}(1-t)e^{i\theta_{sk,f}} + \sigma^{i,f}te^{i\theta_{si,f}}\right) &= 0\\
=\sigma^{k,f}(1-t)\sin(\theta_{sk,f}) + \sigma^{i,f}\cdot t\cdot \sin(\theta_{si,f})\\
\end{align*}
And similarly for $g$. It is not even necessary to control the sign of the real part, since our positivity constraint $\sigma \geq \epsilon$ and correct sum of $\theta$ will guarantee the oppositeness; the only thing that matters is that the collected $\sigma^e$ maintain the proper ratio on each side.

\section{Algorithmic details}
\subsection{Constructing D}
Consider the edge-based (on the beveled mesh) quantity $\beta$, which might be the result of a differential $\beta = d_0\alpha$, but not necessarily, and is represented as a column vector of size $|\mathcal{E}|$. To implement the flap-based finite-volume integral (Eq. 9 in the main paper), we construct a matrix $D:2|\mathcal{E}|\times |\mathcal{E}|$. Consider a single flap with left face $f = kij$, right face $g = ikl$, and edge face on edge $ki$ (like Figure~\ref{fig:edge-singularities}). We choose an arbitrary, but single coordinate system for the flap, so that all edges can be represented as 2D vectors. $e^\perp$ is a rotated edge against the normal of the respective face. The integral produces a vector (intrinsic of size 2) per flap, where the corresponding rows of D would comprise:
\begin{align}
\nonumber D(f \cup g, ij) &= \frac{1}{6}(e_{ki}^\perp-e_{jk}^\perp),\ \ \  \ \ \ 
&D(f \cup g, jk) &= \frac{1}{6}(e_{ij}^\perp-e_{ki}^\perp),\\
\nonumber D(f \cup g, (ki,f)) &= \frac{1}{6}(e_{jk}^\perp-e_{ij}^\perp),\ \ \ \ \ \  
&\nonumber D(f \cup g, kl) &= \frac{1}{6}(e_{ik}^\perp-e_{li}^\perp),\\
\nonumber D(f \cup g, li) &= \frac{1}{6}(e_{kl}^\perp-e_{ik}^\perp),\ \ \ \ \ \ 
&\nonumber D(f \cup g, (ik,g)) &= \frac{1}{6}(e_{li}^\perp-e_{kl}^\perp),\\
\nonumber D(f \cup g, (i, gf)) &= \frac{1}{2}\lambda_{J}e_{ki}^\perp,\ \ \ \ \ \  &D(f \cup g, (k, gf)) &= \frac{1}{2}\lambda_Je_{ki}^\perp.
\end{align}
Which blend the quantities $\beta_{ij},\beta_{jk},\beta_{ki,f},\beta_{ik,g}, \beta_{kl}, \beta_{li}$, and the jump quantities $\beta_{k,fg}, \beta_{i,fg}$.
We assumed this orientation of the edge $1$-forms for clarity and without loss of generality.

\subsection{Alignment}
\begin{figure}
    \includegraphics[width=.45\linewidth]{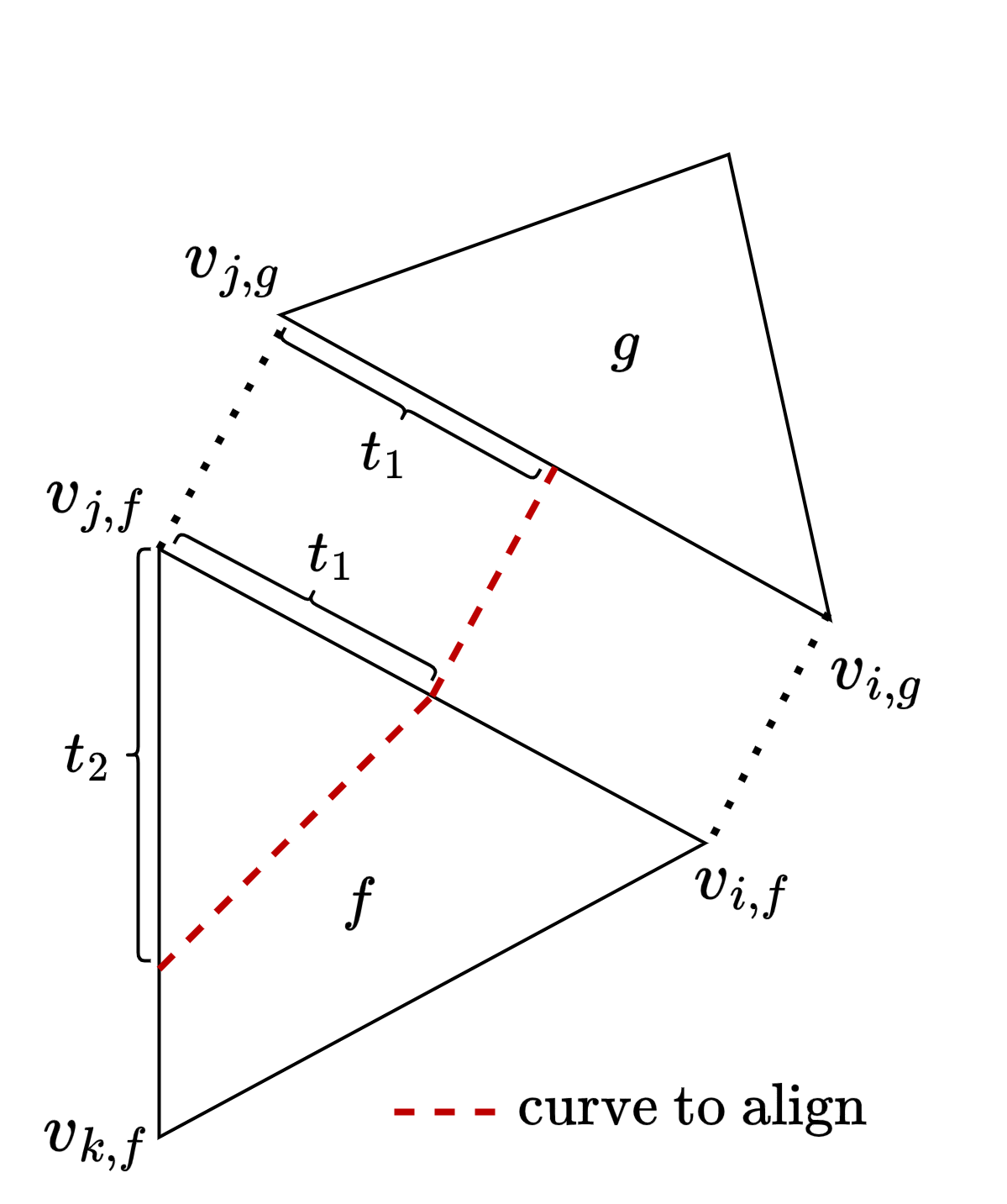}
    \caption{Discrete segments of an alignment curve.}
    \label{fig:alignment-curve}
\end{figure}

Consider a set of alignment curves across the mesh along which the features are to be aligned. The curves are discretized to per-face segments that subdivide the faces of the beveled mesh (Fig.~\ref{fig:alignment-curve}). The essence of our alignment extension is to:
\begin{itemize}
    \item Compute part-edge $\theta$ so that they create zero rotation along the segments, similar to the half-cycles of  Sec.~\ref{subsubsec:edge-singularities}. This results in zero rotations along the alignment curves, but not in global alignment. 
    \item Compute a vertex-based field $\alpha$ so that we augment $\theta$ by $\hat{\theta} = \theta + d_0\alpha$, and where $\alpha$ ``synchronizes''  the alignment between curves (similar to the approach taken in~\cite{Crane:2010}). By definition $\alpha$ does not harm any topological constraints in the rest of our method since $d_1\alpha = 0$. 
\end{itemize}

For simplicity, we just assume the part-edge $\theta$ is initially subdivided linearly per edge, and use $\alpha$ to maintain the correct half-cycle sums in addition to global alignment. This step is added after computing the original $\theta$ and before computing scale with $\hat{\theta}$ replacing $\theta$ (without modifying either steps), and thus it is just a ``plug-in'' to our original algorithm. We next show how to compute $\alpha$ to constrain the tangential alignment of the field to the curve.
\begin{itemize}
    \item \textbf{Original Face crossing.} For a face $f = ijk$ crossed by a curve, we align the features along the segment between the entry and exit points of the curve. Suppose (without loss of generality) the curve first crosses edge $ij$ at fractional length $t_1$, and then edge $jk$ at $t_2$. Assuming piecewise linearity of $\psi$ away from singularities, we impose the constraint:
    $$
    t_1(\alpha_{j, f} - \alpha_{i, f}) + t_2(\alpha_{k, f} - \alpha_{j, f}) = -t_1\theta_{ki} - t_2\theta_{ij, f}.
    $$
    \item \textbf{Edge-Face crossing.} For an edge $ij$ crossed directly, we constrain the jump terms:
    $$
    t_1(\alpha_{j,f} - \alpha_{i,f}) + (\alpha_{j,g} - \alpha_{j,f}) + t_2(\alpha_{i,g} - \alpha_{j,g}) = -t_1\theta_{ij,f} - \theta_{j,fg} - t_2\theta_{ji,g},
    $$
    where $f$ and $g$ are the adjacent faces.
    \item \textbf{Vertex passing.} For a vertex touched by a curve, we impose that all corner-to-corner jumps counter-balance the local phase differences, enforcing
    $$
    \alpha_{i, f} - \alpha_{i, g} = -\theta_{i, fg},
    $$
for each jump edge from $f$ to $g$ at the vertex $i$.
\end{itemize}
Additionally, to align across disjoint curves, we choose one end corner for each curve and find a path connecting each pair using a spanning tree between the corners of the disjoint alignment curves, accumulating the theta and curvature along it (modulo $[-\pi, \pi]$) and making the $\alpha$ difference at the two corners offset it. 
Collecting all these constraints into a matrix $C_{\text{align}}$, we solve for smoothest $\alpha$ via a Laplacian minimization
\[
\begin{aligned}
&\alpha = \text{argmin}\left[ \alpha^T d_0^TWd_0 \alpha\right], \\
\text{s.t.} \quad &  C_{\text{align}} d_0\alpha = -C_{\text{align}}\theta
\end{aligned}
\]
Where $W$ is the cot-weight diagonal matrix. We note that as we treat alignment like edge singularities, we need to use the same scale constraints in the scale system.

\bibliographystyle{ACM-Reference-Format}
\bibliography{bibliography}